\documentclass[11pt]{article}
\usepackage{jcapmod}
\usepackage{booktabs}
\usepackage[english]{babel}
\usepackage{amsmath,amssymb,amsbsy,amstext, amsthm, simplewick}
\usepackage{graphicx}
\usepackage{amsfonts}
\usepackage{amssymb}
\usepackage{upgreek}
 \usepackage{exscale,relsize}
 \usepackage[makeroom]{cancel}
\usepackage{soul}

\RequirePackage{color}

\usepackage{colortbl}
\definecolor{rp}{cmyk}{0.2, 1, 0.6, 0}
\definecolor{green2}{cmyk}{0, 1, 0.5, 0}
\definecolor{lightgreen}{cmyk}{0.2, 0, 0.2, 0.2}
\definecolor{lightgray}{cmyk}{0.1,0.2,0,0.1}
\definecolor{lightgray2}{cmyk}{0.4,0.4,0,0.8}
\definecolor{black}{cmyk}{1.0,1.0,1.0,1.0}

\allowdisplaybreaks[1]

\usepackage{colortbl}
\definecolor{lightgreen}{cmyk}{0.2, 0, 0.2, 0.2}
\definecolor{lightgray}{cmyk}{0.1,0.2,0,0.1}
\definecolor{lightgray2}{cmyk}{0.1,0.1,0,0.1}

\makeatletter
\newlength{\apb@width}
\newcommand{\autoparbox}[2][c]{\settowidth{\apb@width}{#2}\parbox[#1]{\apb@width}{#2}}

\makeatother

\numberwithin{equation}{section}

\def\beq{\begin{equation}}
\def\eeq{\end{equation}}

\def\bea{\begin{eqnarray}}
\def\eea{\end{eqnarray}}

\def\d{{\rm d}}

\def\beq{\begin{equation}}
\def\eeq{\end{equation}}
\def\bea{\begin{eqnarray}}
\def\eea{\end{eqnarray}}

\def\d{{\rm d}}

\def\Mp{M_{\rm pl}}
\def\d{{\rm d}}

\def\0{{\boldsymbol 0}}

\def\M{{\cal M}}
\def\A{{\cal A}}
\def\Z{{\cal Z}}

\DeclareRobustCommand{\SkipTocEntry}[4]{}

\begin{document}

\begin{titlepage}

\setcounter{page}{1} \baselineskip=15.5pt \thispagestyle{empty}

\bigskip\

\vspace{1cm}
\begin{center}

{\fontsize{20}{28}\selectfont  \sffamily \bfseries  Signs of Analyticity in Single-Field Inflation}

\end{center}

\vspace{0.2cm}
\begin{center}
{\fontsize{13}{30}\selectfont  Daniel Baumann,$^{\bigstar}$ Daniel Green,$^{\clubsuit, \blacklozenge}$ Hayden Lee,$^{\bigstar}$ and Rafael A.~Porto$^{\spadesuit}$} 
\end{center}

\begin{center}

\vskip 8pt
\textsl{$^\bigstar$ DAMTP, Cambridge University, Cambridge, CB3 0WA, UK}
\vskip 7pt

\textsl{$^\clubsuit$ Canadian Institute for Theoretical Astrophysics, Toronto, ON M5S 3H8, Canada}
\vskip 7pt

\textsl{$^ \blacklozenge$ Canadian Institute for Advanced Research, Toronto, ON M5G 1Z8, Canada}
\vskip 7pt

\textsl{$^\spadesuit$ ICTP South American Institute for Fundamental Research, 01140-070 Sao Paulo, SP Brazil}\\ 

\end{center}

\vspace{1.2cm}
\hrule \vspace{0.3cm}
\noindent {\sffamily \bfseries Abstract} \\[0.1cm]
The analyticity of response functions and scattering amplitudes implies powerful relations between low-energy observables and the underlying short-distance dynamics. These `IR/UV' relations are rooted in basic physical principles, such as causality and unitarity. In this paper, we seek similar connections in inflation, relating cosmological observations to the physics responsible for the accelerated expansion. We assume that the inflationary theory is Lorentz invariant at short distances, but allow for non-relativistic interactions and a non-trivial speed of propagation at low energies. Focusing on forward scattering, we derive a `sum rule' which equates a combination of low-energy parameters to an integral which is sensitive to the high-energy behavior of the theory.
While for relativistic amplitudes unitarity is sufficient to prove positivity of the sum rule, this is not guaranteed in the non-relativistic case. We discuss the conditions under which positivity still applies, and show that they are satisfied by all known  UV completions  of single-field inflation.  
In that case, we obtain a consistency condition for primordial non-Gaussianity, which constrains the size and the sign of the equilateral four-point function in terms of the amplitude of the three-point function.
The resulting bound rules out about half of the parameter space that is still allowed by current observations. Finding a violation of our consistency condition would 
point towards less conventional theories of inflation, or violations of basic physical principles.

\vskip 10pt
\hrule
\vskip 10pt

\vspace{0.6cm}
 \end{titlepage}

\tableofcontents

\newpage

\flushbottom

\newpage

\section{Introduction}
\label{sec:intro}

Causality is one of the fundamental principles of any physical theory. Requiring the response of a system to be causal connects seemingly different phenomena, such as fluctuations and dissipation, or the speed and the attenuation of light in a medium.  These connections are most manifest in frequency space, where causality is encoded in the analyticity of the response function.  Non-trivial relations between physical observables are then simply a consequence of Cauchy's integral theorem, which relates the real and imaginary parts of the response function, as in the Kramers-Kronig relation.
Similar considerations apply to scattering amplitudes: it is widely believed that (micro)causality is reflected in the analytic properties of the  $S$-matrix.  In this case, Cauchy's theorem provides a link between the low-energy (`IR') limit of the scattering amplitude and its high-energy (`UV') behavior.  In this paper, we use analyticity (causality) to derive analogous relations between cosmological observables and the underlying physics of inflation. 

 \vskip 4pt
Connections between observables at low energies with properties at high energies 
 have been explored before in the context of particle physics and cosmology, e.g.~\cite{weinberg2005theV1,weinberg2005theV2, Colangelo:2000dp, Adams:2006sv,Distler:2006if}.  In particular, the analyticity of scattering amplitudes, together with unitarity and crossing symmetry, has been exploited to derive so-called `sum rules' (or `dispersion relations') relating parameters of the low-energy theory to integrals over 
high-energy cross sections~\cite{weinberg2005theV1,weinberg2005theV2, Colangelo:2000dp}. In some cases, unitarity implies that certain low-energy parameters must be positive~\cite{Adams:2006sv}. (The violation of this positivity condition in the DGP model of modified gravity~\cite{Dvali:2000hr} highlighted that the corresponding theory lacks a local Lorentz-invariant UV completion.) Likewise, in~\cite{Distler:2006if}, this reasoning was applied to the scattering of longitudinal gauge bosons to falsify models of new physics in the electroweak sector.  In the present work, we seek for similar IR/UV connections in single-field inflation. 

 \vskip 4pt 
The IR theory is described in terms of a Goldstone boson of spontaneously broken time translations~\cite{Creminelli:2006xe,Cheung:2007st}, which we denote by the field~$\pi$.  This captures a large class of inflationary models, namely all models of single-field inflation or any model with a single dynamical degree of freedom (or `clock') at horizon crossing.\footnote{Dissipative single-clock models and excited states may also be studied within the EFT framework~\cite{LopezNacir:2011kk,LopezNacir:2012rm,Flauger:2013hra}.}
Moreover, we will assume that the UV theory is Lorentz invariant but allow for a non-relativistic speed of propagation, as well as Lorentz-symmetry breaking interactions, in the effective field theory (EFT) which characterizes the Goldstone dynamics at low energies.\footnote{As we shall see, this introduces extra subtleties in the derivation of the sum rule. For instance, we can have singularities which are not directly associated with propagating degrees of freedom at that scale.}
Interestingly, quantum vacuum fluctuations of a weakly interacting Goldstone boson are sufficient to describe all current observations, without the need to introduce additional light degrees of freedom~\cite{PlanckInflation}.  
However, as measurements become more precise, higher-order Goldstone self-interactions may be detected, or at least will be further constrained. The sum rule that we derive in this paper will be relevant for interpreting future measurements and to test possible deviations from the canonical framework. 
 
 \vskip 4pt
One of the key parameters that can be measured is $c_s$, the speed of propagation of the Goldstone boson. Constraints on primordial non-Gaussianity imply $c_s \geq 0.024\ (95\%\,{\rm CL})$~\cite{PlanckNG}.  Moreover, at leading order in derivatives, and to quartic order in fluctuations, the EFT for $\pi$ contains two additional low-energy parameters, which we will denote by $c_3$ and $c_4$, and which are associated with the interactions $\dot \pi^3$ and $\dot \pi^4$, respectively.  (See~(\ref{equ:Mn}) for the precise definition of the parameters $c_3$ and $c_4$.)

\begin{figure}[h!]
\centering
\hspace{1cm}\includegraphics[scale=1.0]{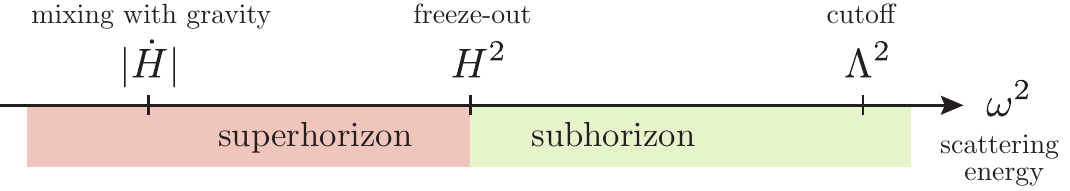}
\caption{Illustration of the relevant energy scales of the EFT. The flat space approximation applies for scattering energies above the Hubble scale, $\omega^2 > H^2$. The decoupling limit captures the regime $\omega^2 > |\dot H|$. The hierarchy $H^2 \ll \Lambda^2$ is guaranteed by the high degree of Gaussianity of the primordial perturbations, while $|\dot H| \ll H^2$ holds as a condition for inflation and is supported by observations of the spectral index.} 
\label{fig:scales}
\end{figure}

We will constrain these three parameters of the EFT by studying $\pi \pi \to \pi \pi$ scattering, as a function of the center-of-mass energy $\omega$.\footnote{For energies below the cutoff of the EFT, $\omega \ll \Lambda$, $S$-matrix elements may be computed using $\pi$ as the interpolating field. Equivalently, we may use the scalar curvature perturbation $\zeta$, which is also guaranteed to be present in the UV.}  To perform the computations, we will exploit the natural hierarchy of scales in the problem (see fig.~\ref{fig:scales}).  
Since Goldstone bosons are derivatively coupled, their scattering amplitudes are dominated by high-energy (short-distance) processes near the cutoff scale $\Lambda$ of the EFT. Moreover, at lower energies, only a handful of terms contribute since higher-order terms are suppressed by inverse powers of the cutoff. These features will allow us to work in the {\it flat space limit} ($H \to 0$) and compute scattering amplitudes without taking into account the cosmological expansion. Corrections to our results will be suppressed by powers of $H^2/\Lambda^2$.
We will also use the so-called {\it decoupling limit} ($\Mp^2 \to\infty$, $\dot H \to 0$, with $\Mp^2 \dot H  = const.$), in which the mixing between $\pi$ and gravitational perturbations vanishes.
Computations performed in the decoupling limit will be accurate up to corrections that scale as $\dot H /H^2$ and $\omega^2/ \Mp^2$. 
A somewhat unusual fact of the flat space and decoupling limits is that slow-roll inflation turns into a free theory --- i.e.~the inflaton potential $V(\phi)$ becomes constant and gravitational interactions are turned off. When these limits are taken, $\pi\pi$ scattering therefore has a trivial scattering amplitude for slow-roll inflation. 

\vskip 4pt
Within these approximations we will derive a sum rule that links the three parameters, ($c_s, \hskip 1pt c_3, \hskip 1pt c_4$), to an integral over the imaginary part of the forward scattering amplitude~$\A(s)$, where $s \equiv 4 \omega^2$ is the square of the energy in the center-of-mass frame. The integral will get contributions from branch cuts or poles arising from the production of intermediate states.
 Along the positive real axis, $s >0$, unitarity of the $S$-matrix enforces positivity of the integral.  On the other hand, positivity on the negative real axis, $s<0$, is not guaranteed for our non-relativistic system. Nevertheless, there exists a scale $\rho$ at which Lorentz symmetry is restored. For~$|s| > \rho^2$,  crossing symmetry then relates $s$- and $u$-channel processes and the contribution to the integral is still positive.
 Below the scale~$\rho$, however, we cannot rule out cuts and poles on the negative axis that give negative contributions. 
 When these terms are present, and dominate, unitarity is not sufficient to enforce that the integral is positive, unlike in relativistic theories. That being said, all known examples of UV completions of the EFT of inflation 
exhibit positivity.  As we will show,  violations may occur only under rather peculiar conditions. It is therefore interesting to investigate the consequences of positivity in the sum rule.

\vskip 4pt
When all contributions to the sum rule are positive, it will enforce the positivity of a certain combination of the EFT parameters ($c_s, \hskip 1pt c_3, \hskip 1pt c_4$). This then leads to a new {\it consistency condition}\footnote{We use the term `consistency condition' in the same fashion as in the single-field consistency condition~\cite{Maldacena:2002vr,Creminelli:2004yq}. Exceptions to our results will point to specific violations of our assumptions, for otherwise consistent theories.} relating the size and the sign of the parameter  $c_4$ to the values of $c_s$ and~$c_3$. 
For the special case in which the interactions are dominated by the parameters $c_3$ and $c_4$ (corresponding to $|c_4| \gtrsim |c_3| \gg 1$), we get:
\beq
\label{1.1}
c_4 \, >\, (2c_3)^2\ .
\eeq
A weaker condition, $c_4 > - 1$, holds for any values of $c_s$ and $c_3$. The most general expression of our bound for arbitrary values of $c_s$, $c_3$, and $c_4$, is given in Section~\ref{sec:application}: cf.~(\ref{equ:c4-2}). 

 \vskip 4pt 
The consistency condition has both observational and theoretical consequences. For example, in~\cite{Senatore:2010jy} it was pointed out that $|c_4| \gg 1$ is technically natural, in the sense of `t Hooft~\cite{natural}, since it corresponds to an emergent $\pi \to -\pi$ symmetry of the Goldstone action. 
However, while both signs of $c_4$ are natural in this sense, \eqref{1.1} can only be satisfied for positive $c_4 \gg 1$.  
This is to be compared with the current observational bound~\cite{PlanckNG}: 
\beq
- 8.3 \times 10^7   \,<\, c_4/c_s^4 \,<\, 7.4 \times 10^7  \quad (95\%\hskip 1pt\rm CL).  
\eeq
We see that about half of the parameter space that is still allowed by observations would be ruled out by our theoretical considerations, provided positivity of the sum rule applies.

\vskip 4pt 
A violation of our consistency condition could arise from negative contributions to the sum rule, although we will argue that this would require less conventional models of inflation.  More drastically, a violation could signal the breakdown of some basic properties of the UV completion of the EFT of inflation, such as causality, unitarity and Lorentz invariance.\footnote{In fact, we will demonstrate that our bound is closely related, but not equivalent, to the requirement that the theory does not allow superluminal propagation around non-trivial backgrounds. See also \cite{Camanho:2014apa}.}  Hence, testing our consistency condition provides very useful information about the physics of inflation.  We believe that this further justifies the continuing experimental effort for improving current bounds on non-Gaussianity, including joint constraints on the primordial three- and four-point functions~\cite{PlanckNG}.

\vskip 4pt
As for the case of light in a medium, one may ultimately hope to connect the value of $c_s$ (or other measurable quantities) to the microphysics underlying the early phase of accelerated expansion. Unfortunately, our study of forward scattering does not lead directly to a sum rule for $c_s$ alone. We will speculate that such
a sum rule may be obtained for non-forward scattering, since the angular dependence of the scattering amplitude depends solely on an interaction proportional to $(1-c_s^2)$, 
or through generalised Kramers-Kronig relations for the Green's function.
The hypothetical form of the sum rule, together with positivity,  motivates a tantalizing conjecture: The only Lorentz-invariant UV completion of a $c_s=1$ theory obeying the basic properties of the $S$-matrix~\cite{smatrix} 
is slow-roll inflation (i.e.~a free theory in the flat space and decoupling limits). 

\vskip 4pt
The outline of the paper is as follows. In Section~\ref{sec:sum}, we review 
the analytic properties of relativistic and non-relativistic scattering amplitudes.
We derive a sum rule which relates the real part of the forward amplitude at low energies to an integral over its imaginary part.
In Section~\ref{sec:application}, we assume positivity of the integral to derive constraints on a combination of the parameters of the EFT of inflation, including a consistency condition relating the quartic and cubic couplings.
Moreover, armed with the full amplitude, we present an improved derivation of the critical sound speed for which the EFT admits a perturbative UV completion~\cite{Baumann:2014cja}.
 In Section~\ref{sec:atwork}, we explicitly demonstrate the validity of the sum rule  for the weakly coupled completion of~\cite{Tolley:2009fg}. We show that the positivity constraints are satisfied, and argue that this is a generic feature of a large class of weakly coupled UV completions of the EFT of inflation. We also provide evidence for the conjecture that $c_s = 1$ is only compatible with slow-roll inflation. We discuss the observational implications of our results in Section~\ref{sec:conclusion}. Technical details are relegated to appendices.

\subsection*{Notation and Conventions}

Our metric signature is ($-$\hskip 1pt$+$\hskip 1pt$+$\hskip 1pt$+$). We will use
natural units, $c=\hbar \equiv 1$, and define the reduced Planck mass as $\Mp \equiv (8\pi G)^{-1/2}$. The letter $\pi$ will refer both to $3.141\ldots$ and the Goldstone boson of broken time translations. We write three-momenta as $\vec{k}_a$ and four-momenta as~$p_a = (\omega_a, \vec{k}_a \hskip 1pt)$, where $a=\{1,\cdots \hskip -1pt,4\}$ labels the momentum of each particle. We also use the traditional Mandelstam variables ($s, \hskip 1pt t,\hskip 1pt u$) for relativistic $2\to2$ scattering. 
We will follow the conventions of~\cite{Peskin} and write the $S$-matrix as 
\beq
\langle p_3p_4 | \, S\, |p_1 p_2 \rangle =(2\pi)^4 \delta(p_1+p_2 -p_3-p_4) \big[1+ i \M(s,\theta)\big] \ ,
\eeq
where $\cos\theta \equiv \hat{k}_1 \cdot \hat{k}_3$ is the scattering angle. 
We denote the amplitude in the forward limit by
\beq
\label{dashA}
\A(s) \equiv \lim_{\theta\to 0}\M(s,\theta)\ .
\eeq 
For non-relativistic scattering, we will find it convenient to introduce a new set of variables.
Since defining $\tilde p_a = (\omega_a, c_s(\omega_a)\hskip 1pt \vec{k}_a)$ restores (a fake) relativistic invariance of the free field part of the action, we will use a set of modified Mandelstam variables in terms of the rescaled momenta:
\beq
\label{tildeeq}
\tilde s \equiv -(\tilde p_1+\tilde p_2)^2\ , \qquad \tilde t \equiv -(\tilde p_1-\tilde p_3)^2\ , \qquad \tilde u = -(\tilde p_1-\tilde p_4)^2\ . 
\eeq

\section{Analyticity and Sum Rules}
\label{sec:sum}

In this section, we will review the standard analyticity arguments for relativistic scattering, see e.g.~\cite{Bella}, and then discuss the additional subtleties that arise if the low-energy limit breaks Lorentz invariance. Some details of the discussion are relegated to Appendix~\ref{sec:analytic}. In Section~\ref{sec:application}, we will apply the formalism to the EFT of inflation.

\subsection{Relativistic Scattering}
\label{sec:rel}

For relativistic interactions, it is natural to consider the amplitude of $2\to 2$ scattering to be a function of the Mandelstam variables $s$ and $t$, i.e. $\M(s,t) \equiv \M(s,\theta(s,t))$.  A minimal amount of non-analytic behavior of $\M(s,t)$ for complex $s$, and at fixed transfer momentum $t$, is required by unitarity of the $S$-matrix: $S S^\dagger = 1$ \cite{smatrix}.  In particular, for forward scattering, $t\to 0$, the {\it optical theorem} allows us to write the imaginary part of the amplitude as
\beq\label{equ:opt2}
\hspace{-1cm} 2\, {\rm Im} [ \A(s)] = \sum_I \int \d \Pi_I \, |\M(p_1,p_2 \to I)|^2 \ ,
\eeq
where $I$ stands collectively for all possible intermediate states, each with a differential phase space element of $\d\Pi_I$.
Using Hermitian analyticity, $\A(s) = \A^*(s^*)$, one may also write
\begin{align}
2i\, {\rm Im} [ \A(s)]  &\equiv \A(s+i\epsilon) - \A^*(s+i\epsilon) \nonumber \\
&=\A(s+i\epsilon) - \A(s-i\epsilon) \equiv {\rm Disc}[\A(s)]\ , \label{disc}
\end{align}
where ${\rm Disc}[\A(s)]$ denotes the discontinuity of $\A(s)$ across the real axis. The hypothesis of maximal analyticity\footnote{This hypothesis can be demonstrated in perturbation theory -- see~\cite{smatrix}. However, one cannot rule out the possibility of non-trivial analytic behavior due to non-perturbative physics (e.g.~\cite{GiPo}).}
then assumes that $\A(s)$ is non-analytic {\it only} when ${\rm Im}[\A(s)] \ne 0$ along the real axis, i.e.~when the right-hand side of (\ref{equ:opt2}) is non-zero above the mass thresholds for the states~$I$.
For the physical domain $s> 0$, this determines the locations of poles and branch cuts in terms of the energies of the states~$I$.   Moreover,  the non-analytic behavior of $\A(s)$ for the unphysical values $s< 0$ is dictated by {\it crossing symmetry}.  Specifically, 
there is a connection between the amplitude at $s+i\epsilon$ (above the branch cuts) and that at $-s-i\epsilon$ (below the branch cuts), which may be shown to exist even for massless particles~\cite{BEG}. For identical particles, this implies that the forward amplitude is an even function, i.e.~$\A(s)= \A(-s)$, and the singularities in the complex $s$-plane can all be accounted for in terms of $s$- and $u$-channel exchanges.

\begin{figure}[h!]
\centering
\hspace{1cm}\includegraphics[scale=0.5]{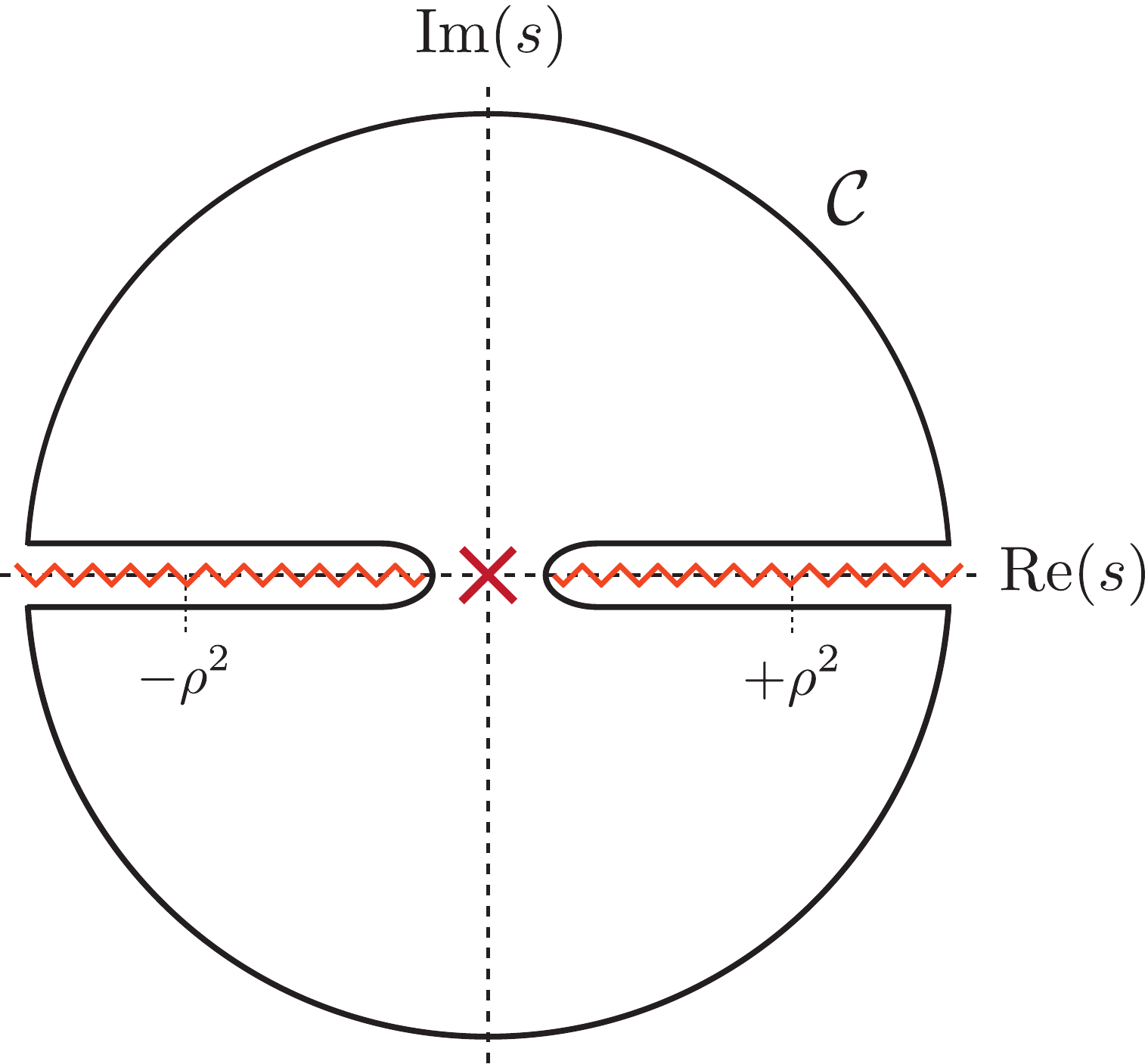}
\caption{Illustration of the choice of contour in~(\ref{equ:A5}).  \label{contours}}
\end{figure}

When the intermediate states  $I$ are massive, there is a gap between the singularities on the real axis. 
Considering the function $\A(s)/s^3$, Cauchy's theorem then implies the following sum rule
\beq
\frac{1}{2}\A^{\prime \prime}(s\to0) = \oint\limits_{\cal C} \frac{\d s}{2\pi i} \frac{\A(s)}{s^3} \ ,  \label{equ:A5}
\eeq
where $\A^{\prime \prime}(s) \equiv \partial_s^2 \A(s)$ and ${\cal C}$ is the contour illustrated in fig.~\ref{contours}.
The Froissart-Martin bound~\cite{Froissart:1961ux,Martin:1962rt},  $|\A(s)| \le const. \times s \ln^2 s$, for $|s| \to \infty$, lets us drop the contour at infinity, and only the discontinuities across the branch cuts, ${\rm Disc}[\A(s)] \equiv 2 i\hskip 1pt {\rm Im}[\A(s)]$, contribute to the right-hand side of (\ref{equ:A5}),
\beq
\A^{\prime \prime}(s\to0) = \frac{2}{\pi}
\left( \int_{-\infty}^0 + \int_0^\infty \right) \d s\, \frac{{\rm Im}[\A(s)]}{s^3}  \ .  \label{equ:A5x}
\eeq
The Froissart-Martin bound may be violated when massless particles are present.\footnote{In general the amplitude remains polynomially bounded on the physical sheet \cite{GiPo}, so that an integral similar to \eqref{equ:A5} can be written for some $n$-th derivative of the amplitude.} However, in our case, the sum rule in \eqref{equ:A5x} still applies. 
As we shall see, this is because the forward scattering limit happens to be free of singularities,\footnote{Let us remark that turning on gravitational interactions unavoidably induces divergences in the forward direction. While this is a general problem --- gravity is universal --- we expect the proper treatment of these effects to be highly suppressed in the cosmological setting, $H/\Mp \ll 1$, and not to modify significantly the results derived from forward scattering. A similar attitude is used to ignore singularities from photon exchanges in QCD processes.} which permits the application of  standard techniques 
to derive the number of necessary subtractions, e.g.~\cite{Jin}.\footnote{Alternatively, we could introduce a small mass and later send it to zero. 
For scalar particles this does not modify the structure of the theory nor the UV behavior, which therefore still obeys the Froissart-Martin~bound.}

Using~(\ref{equ:opt2}) to write the imaginary part of the amplitude in terms of the cross section, i.e.~${\rm Im}[\A(s)] \equiv s\sigma(s)$, and crossing symmetry which relates the integrals on the positive and negative axes, ${\rm Im}[\A(-s)] = - {\rm Im}[\A(s)]$, we get the sum rule in its final form:
\beq
\A^{\prime \prime}(s\to0) \,=\, \frac{4}{\pi} \int_0^\infty  \d s\, \frac{\sigma(s)}{s^2}  \ . \label{equ:App}
\eeq
The right-hand side of (\ref{equ:App}) is manifestly positive, which is a 
consequence of unitarity.
Extensions to non-forward scattering are possible, even for unphysical values of~$t$~\cite{Jin}. However, except for a speculative conjecture in \S\ref{sec:cs=1}, we will concentrate on forward scattering. 

\subsection{Non-Relativistic Scattering}
\label{ssec:NR}

We now consider the extension to non-relativistic scattering.  
We assume that the theory is Lorentz invariant in the UV, but allow for a non-trivial sound speed $c_s \neq 1$, as well as other Lorentz-symmetry breaking interactions, in the IR. For simplicity, we will work in the center-of-mass frame, where the forward amplitude $\A_{\rm cm}$ becomes a function of the square of the center-of-mass energy, $4\omega^2$, which also coincides with the Mandelstam variable $s$ in this particular frame.
To match the low-energy and high-energy behaviors of the scattering amplitude, we write $\A(s) \equiv \A_{\rm cm}(4\omega^2)$.
 We suppress the `cm' subscript from now on. However, as we describe in Appendix~\ref{sec:analytic},  away from the center-of-mass frame, the forward scattering amplitude in non-relativistic theories is typically not a function of only the Mandelstam variable $s$.

\vskip 4pt 
 The argument for analyticity of the scattering amplitude off the real axis is similar to the relativistic case. However, for non-relativistic theories, the amplitude ${\cal A}(s)$ is not guaranteed to be symmetric under $s \to - s$. Hence, the sum rule~\eqref{equ:A5x} still applies, but the relationship between the contributions for positive and negative $s$ needs to be reconsidered.  In particular, the behavior for $s<0$ is not directly determined (via crossing symmetry) by that at $s>0$. We discuss the subtleties of the non-relativistic case in detail in Appendix~\ref{sec:analytic} and illustrate the novel features in a specific example in Appendix~\ref{sec:example}. Here, we just summarize the main results.

\vskip 4pt
We will assume the existence of a high-energy scale, $\rho$, above which the theory becomes relativistic. As a consequence, $\A(s)$ satisfies the relativistic crossing symmetry for $|s| \gg \rho^2$. The contribution to the integral in \eqref{equ:A5x} from $s \in (-\infty, -\rho^2]$ can therefore be mapped to $s \in [+\rho^2,  +\infty)$, and we can write the sum rule as\footnote{As we emphasized in the introduction, we will work in the flat space limit, which will allow us to apply this expression to the EFT of inflation. 
This is why we have taken $s\to 0$ 
on the left-hand side of (\ref{equ:disperA}), and also in the limits of the integrals on the right-hand side. 
In a derivative expansion, the error induced in the left-hand side will be of order $H^2/\Lambda^2$, where $\Lambda$ is the cutoff of the EFT. (This uses the fact that Goldstone bosons are only derivatively coupled.) The flat space approximation is more accurate on the right-hand side of (\ref{equ:disperA}). This is because the part of the integral which can be computed within the EFT is dominated by short-distance processes near the cutoff $\Lambda \gg H$. In addition, the rest of the integral includes contributions from higher energies, $\omega \gg \Lambda$, where the effects of the cosmological expansion are even less relevant.}
\beq\label{equ:disperA}
\A^{\prime \prime}(s\to0) \,=\, \underbrace{\frac{2}{\pi} \left( \int_{0}^{\infty} + \int_{\rho^2}^\infty  \right) \d s\, \frac{{\rm Im}[\A(s)]}{s^3} }_{> 0} \ \ +\ \ \underbrace{\frac{2}{\pi} \int_{-\rho^2}^0 \d s\, \frac{{\rm Im}[\A(s)]}{s^3}}_{?} \ \ .
\eeq 
The integral above $\rho^2$ is positive definite, since it corresponds to the cross section to produce high-energy states in the theory, which we assume is dominated by relativistic interactions.\footnote{As we shall see, the forward scattering amplitude in the EFT is dominated by contact terms without long-range interactions, and therefore high energies are directly connected with short distances. In general, high-energy (e.g. super-Planckian) exchanges may still remain in a non-relativistic regime for very large impact parameters~\cite{nrgr1,nrgr2}.}
 
Furthermore, for derivatively-coupled theories like the EFT we study in Section~\ref{sec:application}, the leading order amplitude at low energies, $0<s \ll \rho^2$, is analytic in $s$. This is because particle production will be suppressed by extra factors of $s$ over the cutoff scale of the EFT.\footnote{As we will discuss, the cutoff scale of the EFT, $\Lambda$, may be different from $\rho$.} In other words, the tree level contribution dominates the amplitude since $\A (s) \propto s^2$ and ${\rm Im}[\A] \propto |\A|^2 \propto s^4$. Therefore, at leading order, the branch cuts induced by loops of light particles do not contribute to ${\rm Im}[\A]$. As we will see, other singularities for $0 < s < \rho^2$ (e.g.~poles) do not appear unless extra light degrees of freedom are present. We therefore conclude that the first term on the right-hand side of (\ref{equ:disperA}) is manifestly positive. Only the region~$-\rho^2 < s< 0$ may potentially lead to a negative contribution to the sum rule. In \S\ref{sec:WeakExample}, we study an explicit example, in which an extra pole appears in the region $-\rho^2 \ll  s < 0$. 
Nevertheless, positivity is still preserved in this example, and more generally for a large class of weakly coupled completions of small-$c_s$ theories. In general, violations of positivity require large contributions from the $u$-channel whose signs are not fixed by the equivalent $s$-channel exchange. Although we cannot rule out such exotic (plausibly strongly coupled) possibilities, we are yet to encounter an explicit example. Nonetheless, we will take an agnostic attitude towards positivity, and in Section~\ref{sec:application} we will derive constraints on the EFT of inflation by assuming a positive right-hand side of the sum rule.  We believe these to be valuable consistency conditions on a vast class of single-field models with Lorentz-invariant UV completions. The same way a violation of the consistency condition derived in \cite{Maldacena:2002vr,Creminelli:2004yq} would require us to abandon the single-field hypothesis, violations of the positivity constraints we find here, although unlikely, would require us to incorporate the rather peculiar behavior we have identified  
in a full theory of inflation.

\section{Implications for the EFT of Inflation}
\label{sec:application}

In this section, we show how analyticity and unitarity of $\pi \pi  \hskip -1pt  \to \hskip -1pt \pi \pi$  scattering constrains the parameters of the EFT of inflation.
In \S\ref{ssec:Lagrangian}, we present the effective Lagrangian for the Goldstone boson~$\pi$, at leading order in derivatives and to quartic order in fluctuations. 
We use this Lagrangian,\footnote{Following~\cite{LopezNacir:2011kk,LopezNacir:2012rm,Flauger:2013hra}, it is in principle possible to extend our analysis to dissipative single-clock models or theories with excited initial states. However, extra care is required when computing scattering amplitudes for particle excitations in non-vacuum states. We leave this for future work.}  in \S\ref{ssec:positivity}, to compute the low-energy limit of the scattering of $\pi$-particles, and derive a positivity bound on the EFT parameters.
In \S\ref{ssec:unitarity}, we discuss perturbative unitarity of the scattering amplitude, in terms of its partial wave decomposition. 
We show that d-wave scattering leads to an improved derivation of the critical sound speed for which the EFT admits a perturbative UV completion~\cite{Baumann:2014cja}.

\subsection{Goldstone Dynamics}
\label{ssec:Lagrangian}

Let us summarize the basic elements of the EFT of inflation that will be relevant for our discussion.  For more details, we refer the reader to the original papers~\cite{Creminelli:2006xe,Cheung:2007st} or the recent reviews~\cite{Piazza:2013coa, Baumann:2014nda}.

\vskip 4pt
The EFT of inflation is an effective theory of the Goldstone boson, $\pi$, associated with the breaking of time translations in a quasi-de Sitter background.
It parameterizes the low-energy dynamics of adiabatic perturbations in a large class of inflationary models. 
The most general action for $\pi$ to lowest order in derivatives (per field) is~\cite{Cheung:2007st}
\begin{align}
S \ =\ \int \d^4 x \sqrt{-g} \Bigg[ &\ \frac{1}{2} \Mp^2 R + \Mp^2 \dot Hg^{\mu \nu}\partial_\mu (t+\pi )\partial_\nu (t+\pi) - \Mp^2 (3 H^2+ \dot H)   \nonumber \\
&\ \ +  \ \ \sum_n \frac{M_n^4}{n!} \Big(g^{\mu \nu} \partial_\mu (t+\pi )\partial_\nu (t+\pi) + 1\Big)^n \ +\ \cdots \Bigg]\ , \label{equ:S}
\end{align}
where $H(t)$ is the Hubble expansion rate of the inflationary background and $M_n(t)$ are parameters defining the higher-order interactions of the EFT.  The effective action, in principle, includes higher-derivative terms which we did not display~\cite{Cheung:2007st}. However, in the flat space and decoupling limit,\footnote{We assume that the parameters in the EFT are kept fixed as we take $M_{\rm pl}^2 \to \infty$ in the decoupling limit.  For the case of non-zero $M_n^4$, this assumption is necessary for self-consistency.  However, it clearly forbids a non-trivial tensor sound speed, $c_t$. When $c_t \neq 1$ the decoupling limit becomes more subtle, e.g.~\cite{Creminelli:2014wna}. We will explore this possibility elsewhere.} these terms are subdominant at low energies, $\omega/\Lambda \ll 1$, and will not contribute significantly to the left-hand side of the sum rule we derive in this paper.\footnote{Higher-derivative terms may be relevant at low (but finite) energies, provided $c_s \ll 1$, e.g.~\cite{Senatore:2004rj}. The dispersion relation could then be quadratic (or higher order) in $k$ at horizon crossing $\omega \simeq H$. In this case, the flat space approximation becomes more subtle, and we need to account for the scaling of different terms in the action. It is also possible to take a degenerate limit $\Mp^2\dot H \to 0$, which violates the null energy condition. Notice, however, that such specific scenarios are very distinctive from the observational point of view, since small $c_s$ produces a large three-point function \cite{Cheung:2007st}. We will not consider these possibilities here, and assume that theories of inflation have a well-defined flat space limit, for which our bounds apply regardless of the value of $H$. It should be clear, nonetheless, how to modify our analysis to include such cases.}
We will be also interested in the case where $\dot H(t) = \dot H$ and $M_n^4(t) = M_n^4$ are independent of time.  This captures the behavior of the EFT of inflation in the limit of exact scale invariance. Deviations from scale-invariance can be treated perturbatively, but are not relevant for the present work since they are required to be small by measurements of the spectral index.

To relate the low-energy limit of the theory to its high-energy behavior, we will consider the scattering of $\pi$-particles. We will work also in the decoupling and flat space limit(s), as we discussed in the introduction. 
Expanding (\ref{equ:S}) up to quartic order in powers of~$\pi$, we get
\begin{align}
{\cal L}_2 &= \Mp^2 |\dot H| \Big(\dot \pi^2 - (\nabla \pi)^2 \Big) + 2M_2^4\hskip 2pt\dot{\pi}^2\ ,\label{L2}\\[6pt]
{\cal L}_3 &=  \left(2M_2^4 - \frac{4}{3}M_3^4 \right) \dot{\pi}^3 - 2M_2^4 \hskip 2pt\dot{\pi}(\nabla\pi)^2\ ,\label{L3}\\[6pt]
{\cal L}_4 &= \left(\frac{1}{2}M_2^4 - 2M_3^4 + \frac{2}{3}M_4^4\right) \dot \pi^4  - \Big(M_2^2 - 2 M_3^4 \Big) \dot \pi^2 (\nabla \pi)^2 + \frac{1}{2}M_2^4 (\nabla \pi)^4\ ,\label{L4}
\end{align}
where $(\nabla \pi)^2 \equiv \delta^{ij}\partial_i \pi \partial_j \pi$. If $M_2 \ne 0$, then the Goldstone mode propagates with a non-trivial sound speed
\beq
c_s^2 \equiv \frac{\Mp^2|\dot{H}|}{\Mp^2|\dot{H}|+2M_2^4}\ .
\eeq
Sometimes it will be convenient to rescale the spatial coordinate as $\tilde{x}^i= x^i/c_s$, so that (fake) Lorentz invariance is restored in the quadratic part of the action 
\beq
\tilde {\cal L}_2 \equiv c_s^3 {\cal L}_2 = - \frac{f_\pi^4}{2} (\tilde \partial \pi)^2 \ ,
\eeq
where $(\tilde \partial \pi)^2 \equiv g^{\mu\nu}\tilde \partial_\mu \pi \tilde \partial_\nu \pi$ and $f_\pi^4 \equiv 2 \Mp^2 |\dot H| c_s$. The scale $f_\pi$ determines the energy scale of the symmetry breaking and normalizes the amplitude of the power spectrum of $\pi$-fluctuations. The observed amplitude of curvature perturbations, $\Delta_\zeta^2 = (2.142\pm0.049) \times 10^{-9}$~\cite{PlanckParameters}, is reproduced for $f_\pi = (58.64 \pm 0.33)\, H$.
We will find it convenient to normalize the EFT parameters $M_n$ relative to $f_\pi$:
\beq
M_n^4 \equiv c_n\, \frac{f_\pi^4}{c_s^{2n-1}} \ , \label{equ:Mn}
\eeq
where $c_2 \equiv \frac{1}{4} (1-c_s^2)$. The factors of $c_s$ in (\ref{equ:Mn}) ensure that  $c_n \sim {\cal O}(1)$ are natural parameter values even for $c_s \ll 1$. For instance, in DBI inflation~\cite{Silverstein:2003hf}, all $c_n$ are determined by $c_s$ alone; in particular,~$c_3 = -6 c_2^2$ and $c_4 = 60 c_2^3$.
Observational constraints on the parameters $(c_s,c_3,c_4)$ will be presented in Section~\ref{sec:conclusion}.  In the following, we will be concerned with theoretical bounds. 

\vskip 4pt
It will be convenient to write the effective Lagrangian in terms of the canonically normalized field $\pi_c \equiv f_\pi^2 \pi$:
\begin{align}
\tilde {\cal L} &=  - \frac{1}{2}(\tilde \partial \pi_c)^2+ \frac{1}{\Lambda^2} \left[ \alpha_1  \hskip 1pt \dot{\pi}^3_c - \alpha_2 \hskip 1pt \dot \pi_c(\tilde \partial\pi_c)^2 \right] + \frac{1}{\Lambda^4}\left[\beta_1 \hskip 1pt \dot \pi_c^4  -  \beta_2 \hskip 1pt \dot \pi_c^2 (\tilde \partial \pi_c)^2 +\beta_3 (\tilde \partial \pi_c)^4  \right]\ , \label{equ:L4}
\end{align}
where we have introduced the cutoff scale $\Lambda \equiv f_\pi c_s$ and defined the following auxiliary parameters
\begin{align}
\alpha_1 &\equiv - 2 c_2(1-c_s^2) - \frac{4}{3}c_3 \ , \quad \alpha_2 \equiv  2 c_2 \ , \label{equ:alphas}\\[4pt]
\beta_1 &\equiv  \frac{1}{2}c_2 (1-c_s^2)^2 + 2c_3 (1-c_s^2) + \frac{2}{3}c_4 \ , \quad \beta_2 \equiv - c_2(1-c_s^2)- 2c_3 \ , \quad \beta_3 \equiv \frac{1}{2}c_2\ . \label{equ:betas}
\end{align}
The organization of the effective Lagrangian (\ref{equ:L4}) is somewhat unconventional: we have written all interactions in terms of  the `relativistic invariant' $(\tilde \partial \pi_c)^2$ and pure time derivatives $\dot \pi_c$.  This is motivated by the analytic structure of scattering amplitudes, as discussed in Appendix~\ref{sec:analytic}.  The key point is that the `relativistic' part of the interactions will manifestly behave like a Lorentz-invariant amplitude, so we can trace all the subtleties of working in a non-Lorentz-invariant theory to the pure time derivatives.

\subsection{Bounds from Positivity}
\label{ssec:positivity}

In what follows, we will derive a number of constraints on the Lagrangian parameters $c_{n}$ (or equivalently $M_{n}$) from the requirements of analyticity and unitarity of $\pi \pi  \hskip -2pt  \to \hskip -2pt \pi \pi$ scattering.  
Details of the computations are given in Appendix~\ref{sec:computation}.

\vskip 4pt
To gain intuition for the origin of the bounds, we first consider the special case $|c_4| \gg |c_3| \gg 1$. 
In~\cite{Senatore:2010jy}, it was shown that this parameter regime is technically natural, so it is of a particular observational relevance.
In this limit, the cubic Lagrangian is dominated by the $\dot \pi^3$ interaction (since $|\alpha_1| \to \frac{4}{3} |c_3| \gg \alpha_2$), and the quartic Lagrangian is domination by $\dot \pi^4$ (since $|\beta_1| \to \frac{2}{3}|c_4| \gg |\beta_2| \gg |\beta_3|$).
The effective Lagrangian~\eqref{equ:L4} then reduces to 
\begin{align}
\tilde {\cal L} &\,\to\,  - \frac{1}{2}(\tilde \partial \pi_c)^2 - \frac{4}{3}  \frac{c_3}{\Lambda^2} \hskip 1pt \dot{\pi}^3_c + \frac{2}{3} \frac{c_4}{\Lambda^4} \hskip 1pt  \dot \pi_c^4\ .
\end{align}
Computing the forward scattering amplitude in the center-of-mass frame, we find
\begin{align}
{\cal A}(s) &=  \Big( c_4 - (2c_3)^2 \Big) \frac{s^2}{\Lambda^4}\ ,
\end{align}
and positivity, ${\cal A}^{\prime \prime} > 0$,  implies
\beq
\label{c4c3}
\boxed{c_4 > (2c_3)^2}\ , \quad \text{for $|c_4| \gg |c_3| \gg 1$.}
\eeq
We see that positivity simply requires that the contribution from the contact diagram ($\propto c_4$) dominates over that from the exchange diagram ($\propto c_3^2$).
While either sign of $c_4$ is consistent with naturalness, only positive values satisfy the bound (\ref{c4c3}).

\vskip 4pt
It is straightforward to repeat the analysis for the complete Lagrangian \eqref{equ:L4}, i.e.~without taking a special limit of the EFT parameters.
From the cubic interactions, we get  
\begin{align}
{\cal M}_{\dot{\pi}^3} &=  -\frac{9}{4} \alpha_1^2\, \frac{ s^2}{\Lambda^4}\ , \quad
{\cal M}_{\dot{\pi}(\partial\pi)^2} = -4\hskip 1pt \alpha_2^2\, \frac{s^2}{\Lambda^4} \ , \quad
{\cal M}_{\dot{\pi}(\partial\pi)^2\times\dot{\pi}^3} = - 6\hskip 1pt  \alpha_1 \alpha_2\, \frac{ s^2}{\Lambda^4}\ , \label{equ:cubeamp}
\end{align}
while the quartic interactions lead to
\begin{align}
{\cal M}_{\dot{\pi}^4} &= \frac{3}{2} \beta_1\, \frac{ s^2}{\Lambda^4}\ , \quad 
{\cal M}_{\dot{\pi}^2(\partial\pi)^2} = 2  \beta_2\, \frac{ s^2}{\Lambda^4}\ ,\quad
{\cal M}_{(\partial\pi)^4} =   \beta_3\, (3+\cos^2\theta) \frac{s^2}{\Lambda^4}\ . \label{equ:quadamp}
\end{align}
Despite the fact that we have included diagrams that exchange massless particles, we see that the tree level amplitudes are analytic in $s$.  Since these are the lowest dimension operators we could add to the EFT of inflation, we know that any non-analytic behavior in the low-energy limit must enter at higher order in $s$.
Notice also that the amplitude ${\cal M}(s,\theta\hskip 1pt)$ has no divergences as $\theta \to 0$ (due to the derivative nature of the Goldstone interactions) and therefore has a well-defined forward limit:
\begin{align}
{\cal A}(s) &= \sum_N {\cal M}_N(s,0) \nonumber \\[4pt]
&= \left(-\frac{9}{4} \alpha_1^2 -4\alpha_2^2 - 6 \alpha_1 \alpha_2 + \frac{3}{2} \beta_1+ 2 \beta_2 + 4 \beta_3\right) \frac{s^2}{\Lambda^4} \nonumber \\[4pt]
&= \left( c_4+1 - \Big( (2 c_3 + 1) - a(c_s) \Big)^2 - b(c_s) \right) \frac{s^2}{\Lambda^4} \ , \label{equ:As}
\end{align}
where we defined
\beq
a(c_s) \equiv \frac{1-c_s^2}{4}(4+3c_s^2)\ , \quad \ \ b(c_s) \equiv \frac{1-c_s^2}{16}(14+19c_s^2+15 c_s^4) \ . 
\eeq
Positivity now implies that
\beq
\boxed{c_4 + 1 \, >\, \Big( (2 c_3 + 1) - a(c_s) \Big)^2 + b(c_s)}\ . \label{equ:c4-2}
\eeq
Notice that $b(c_s) \ge 0$, for all $c_s \in [0,1]$. The right-hand side of \eqref{equ:c4-2} is therefore positive semi-definite and we conclude that
\beq
\ \ \ \boxed{c_4 + 1 > 0} \ , \quad \text{for \,{\it all}\, values of $c_3$ and $c_s$.}
\eeq
Moreover, in the limit $c_s \to 1$, \eqref{equ:c4-2} becomes
\beq
\label{boundcs1}
\hspace{-1cm}\ \ \boxed{c_4 + 1\,> \, (2c_3 + 1)^2} \ , \quad \text{for $c_s=1$.}
\eeq
As we will see in $\S\ref{sec:cs=1}$, the last constraint can be reproduced by requiring the absence of superluminality around non-trivial backgrounds (with the additional requirement that $c_3 =0$).

\subsection{Perturbative Unitarity}
\label{ssec:unitarity}

Given the full amplitude, $\M(s,\theta)$, we can learn more about the possible UV completions of the EFT by considering the perturbative unitarity of the partial wave amplitudes~\cite{Baumann:2014cja}. Perturbative unitarity will determine the scale at which the EFT becomes strongly coupled, and therefore sets an upper limit on the scale at which new physics must enter in a weakly coupled theory.  These constraints are qualitatively different from the constraints from analyticity which must be satisfied at $s=0$ for self-consistancy of the EFT.  In contrast, perturbative unitarity constrains the extrapolation of the EFT to higher energies from the growth of the amplitude with $s$.  

\vskip 4pt
For this purpose, we write the amplitude in the following form
\beq
\M(s,\theta) = \left[f(c_s,c_3,c_4) + \frac{1-c_s^2}{12} P_2(\cos\theta)  \right] \frac{s^2}{\Lambda^4} \ \equiv\  16\pi \sum_\ell (2\ell+1) a_\ell(s) P_\ell(\cos \theta)\ . \label{equ:Mst}
\eeq
Unitarity of the $S$-matrix requires that ${\rm Im}[a_\ell]=|a_\ell|^2$, which is only consistent if $|{\rm Re}[a_\ell]| < \frac{1}{2}$.  When the tree level amplitude violates this condition, it means that loop corrections must be large and hence the theory is strongly coupled.  
We say that the theory violates `perturbative unitarity'.
Since the amplitude is a function of energy, this determines the energy scale at which perturbation theory breaks down.
For s-wave scattering, $|{\rm Re}[a_0]| < \frac{1}{2}$ can by achieved at all energies, by tuning the parameters in the function $f(c_s,c_3,c_4)$. However, d-wave scattering only involves the sound speed as a parameter and $|{\rm Re}[a_2]| < \frac{1}{2}$ implies
\beq
\frac{1}{60\pi} \frac{1-c_s^2}{c_s^4} \, \frac{\omega^4}{f_\pi^4} \ < \ \frac{1}{2} \ .
\eeq
For a given value of $c_s$, perturbative unitarity will be violated at a specific energy $\omega_\star(c_s)$.
Conversely, requiring the theory to be weakly coupled up to the symmetry breaking scale $f_\pi$, leads to a critical value of the sound speed
\beq
(c_s)_\star = 0.31\ . \label{equ:cs}
\eeq
For $c_s < (c_s)_\star$ the EFT becomes strongly coupled below the symmetry breaking scale.  In other words, weakly coupled theories cannot produce $c_s \leq (c_s)_\star$ without the appearance of additional degrees of freedom {\it below} $f_\pi$. New physics of this type cannot occur in 
slow-roll inflationary models, which thus would be ruled out by a detection of $c_s < c_\star$.\footnote{The current bound $c_s \geq 0.024\ (95\%\,{\rm CL})$~\cite{PlanckNG} still allows for either new (weakly coupled) physics or non-perturbative effects below (or at) $f_\pi$. This is similar to the situation in the pre-LHC/pre-Higgs era in particle physics. For further discussion see~\cite{Baumann:2014cja}.} Notice that, while our conclusions do not rely on the specific value of $c_\star$, the one in~(\ref{equ:cs}) is somewhat smaller than the value found in~\cite{Baumann:2014cja}, $(c_s)_\star =0.47$. The latter was derived from a partial answer to the s-wave amplitude, with $c_3=c_4=0$. Unlike our previous result, the critical value reported here in \eqref{equ:cs} is more robust, and can only be modified by contributions that are higher order in $\omega$.


\section{Sum Rule and Positivity at Work}
\label{sec:atwork}

The sum rule and positivity bounds discussed in the previous section are very general, but also quite abstract.  At the same time, many aspects of scattering are subtle and counterintuitive in the non-relativistic context.  Nevertheless, we have succeeded in deriving a sum rule relating the IR parameters of the EFT of inflation to the high-energy scattering amplitude 
\beq\label{equ:sumrule4}
 \boxed{\frac{1}{\Lambda^4} \left( c_4+1 - \Big( (2 c_3 + 1) - a(c_s) \Big)^2 - b(c_s) \right) \ =\ \frac{1}{\pi} \int_{-\infty}^{\infty} \d s\, \frac{{\rm Im}[\A(s)]}{s^3}  } \ .
\eeq
A further understanding of the physical connection between the low-energy and high-energy behaviors will require a more intuitive understanding of ${\rm Im}[\A(s)]$ in realistic theories.
In this section, we will therefore study specific examples of models that UV-complete the $c_s \ll 1$ and $c_s =1$ limits of the EFT.  We will find that all examples are consistent with our positivity constraints.  When $c_s \ll1$, we will also see how the sum rule works explicitly.  For $c_s =1$, we will find that the positivity constraints are weaker than those derived from requiring subluminality around non-trivial background. We suggest that looking at non-forward scattering would lead to stronger constraints.
Based on those considerations, we will conjecture that $c_s=1$ is always UV-completed by slow-roll inflation --- i.e.~a free scalar field in the flat space and decoupling limits.

\subsection[Perturbative Example with $c_s \ll 1$]{Perturbative Example with $\boldsymbol{c_s \ll 1}$}
\label{sec:WeakExample}

The canonical example of inflation with a small sound speed is DBI inflation~\cite{Silverstein:2003hf,Alishahiha:2004eh}.  While it is easy to show that the positivity bound (\ref{equ:c4-2}) is satisfied for DBI inflation, it is less straightforward to study the high-energy scattering in this theory.  To gain more intuition for how our sum rule works and how positivity arises, it will be instructive to study an example that remains perturbative up to high energies, $\omega \gg f_\pi$.  

\subsection*{The $\boldsymbol{\pi\sigma}$-model}

A reduced sound speed arises for fluctuations around curved trajectories in higher-dimensional field spaces.
A simple two-field model that describes such dynamics is~\cite{Tolley:2009fg} (see also~\cite{Baumann:2011su, Chen:2009zp, Achucarro:2010jv, Achucarro:2012sm, Achucarro:2012yr, Gwyn:2012mw, Cespedes:2012hu, Avgoustidis:2012yc, Chen:2012ge, Assassi:2013gxa, Achucarro:2015bra}):
\beq
{\cal L} \,=\, - \frac{1}{2}k(\sigma)(\partial\phi)^2 - \frac{1}{2}(\partial \sigma)^2 -V(\sigma) \ , \label{equ:PhiSigma}
\eeq
where
\begin{align}
k(\sigma) &\equiv 1 + \frac{\sigma}{M} + \cdots\ , 
\qquad
V(\sigma) \equiv \frac{1}{2}m^2 \sigma^2 + \frac{1}{3!} \mu \sigma^3 + \cdots  \ .
\end{align}
We have suppressed additional terms in the potential for $\sigma$ which stabilize the second field at $\sigma_0 \ll M$; see~\cite{Chen:2009zp, Assassi:2013gxa}. 
The Lagrangian in~(\ref{equ:PhiSigma}) is itself only an EFT, valid at first order in a derivative expansion and up to energies of order $M$. The scale $M$ thus becomes the new cutoff of the theory, which allows for perturbative control provided $ \omega^2 < M^2$.  

\begin{figure}[t!]
\centering
\hspace{1cm}\includegraphics[scale=0.6]{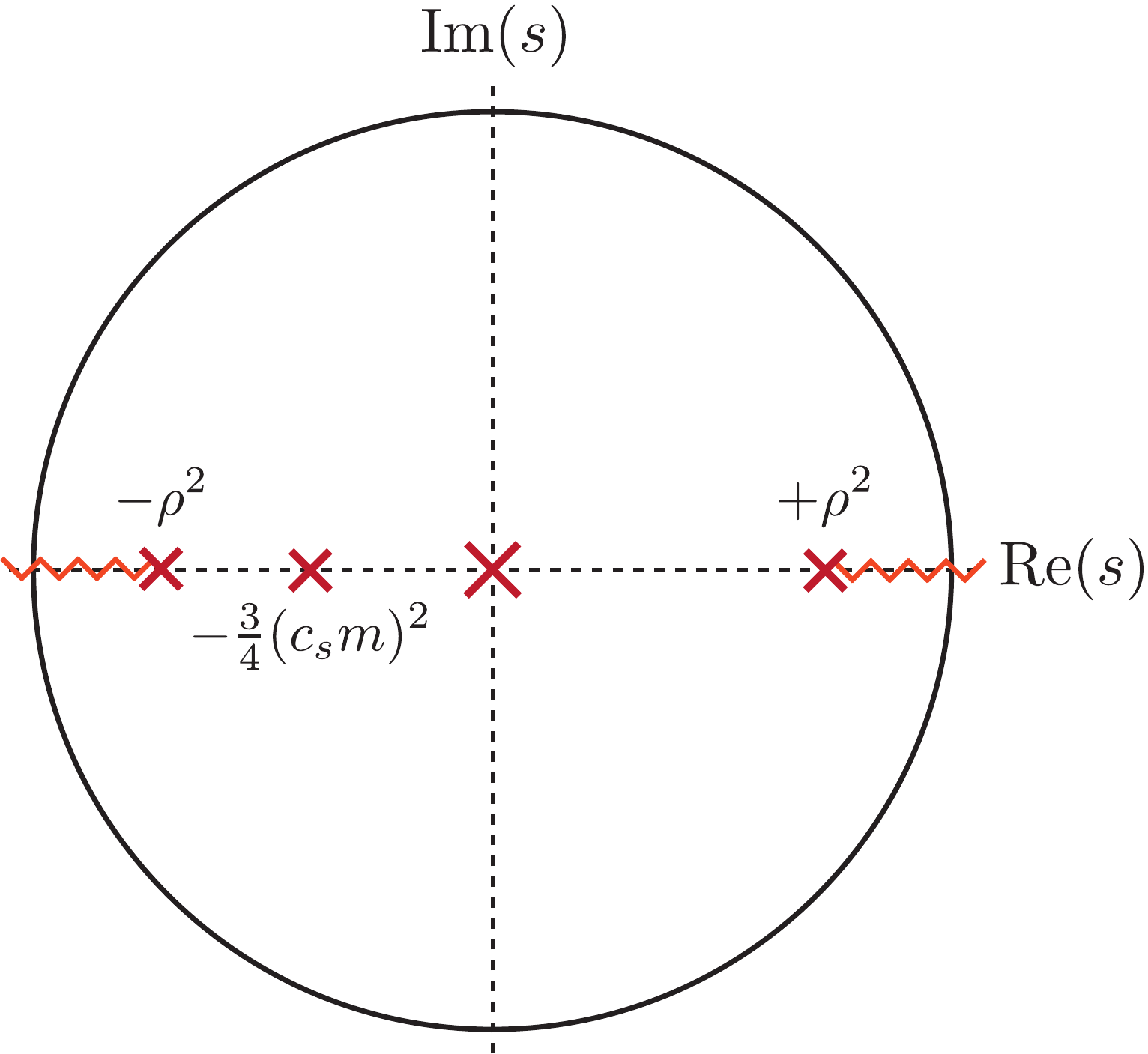}
\caption{Illustration of the pole structure of the amplitude in~(\ref{equ:egamp1}).} \label{contours2}
\end{figure}

\vskip 4pt
Perturbing around the background solution $\phi_0(t)$, i.e.~writing $\phi(t,\vec{x}) \equiv \phi_0(t) + \dot \phi_0 \hskip 1pt \pi(t,\vec{x})$, we get a Lagrangian for the Goldstone fluctuations~$\pi$, coupled to the additional field $\sigma$:
\begin{align}
\nonumber
{\cal L} &\,=\, - \frac{1}{2} |\dot \phi_0|^2 \left[1+\frac{\sigma}{M}\right] \left[-2\dot{\pi}+(\partial\pi)^2\right]  -\frac{1}{2} (\partial \sigma)^2  - \frac{1}{2}m^2 \sigma^2 -\frac{1}{3!}\mu \sigma^3   \ , \\[4pt]
&\,=\, - \frac{1}{2}(\partial  {\bar \pi})^2 -\frac{1}{2} (\partial \sigma)^2    - \rho\hskip 1pt \sigma \dot {\bar \pi} - \frac{\sigma (\partial {\bar \pi})^2}{2M} - \frac{1}{2}m^2 \sigma^2  -\frac{1}{3!}\mu \sigma^3   \ , \label{actexp}
\end{align}
where we have only kept the leading order terms.
In the second line, we have defined
 $ {\bar \pi} = |\dot \phi_0| \pi$ and $\rho \equiv  |\dot \phi_0|/ M$.  
In the following, we will assume the hierarchy of scales 
\beq 
\mu^2 \lesssim m^2 \ll \rho^2\ .
\eeq
The dynamics of the  Lagrangian~(\ref{actexp}) are discussed in detail in~\cite{Baumann:2011su, Assassi:2013gxa}.
At high energies, $\omega > \rho$, the theory describes two relativistic scalars, whose interaction can be treated as a small perturbation.
Below $\omega = \rho$, the mixing becomes relevant and the theory reduces to a single propagating degree of freedom. 
For $m < k < \rho$, the dispersion relation of the Goldstone $\pi$ is nonlinear, $\omega = k^2/\rho$. 
As explained in~\cite{Baumann:2011su}, integrating out the field $\sigma$ produces a non-local action for $\pi$, which is not captured by (\ref{equ:S}). In order to have a local description requires keeping the field $\sigma$, even though it then plays the role of an auxiliary field.\footnote{Let us emphasize that most of these features appear because Lorentz invariance is spontaneously broken, and are commonplace, for example, in non-relativistic condensed matter systems.} For $k<m$ (or $\omega \lesssim c_s m$), the dispersion relation becomes linear, and the low-energy EFT is characterized by a reduced sound speed
\beq 
c_s^2 = \frac{m^2}{m^2+\rho^2} \simeq \frac{m^2}{\rho^2}\ . \label{cseq}
\eeq   
The effective theory is thus described in terms of \eqref{equ:S} without reference to $\sigma$. 
Notice that, for $c_s \ll 1$, the range of validity of the single-field EFT description is  smaller than the naive expectation, which associates the cutoff of the EFT with the mass of the particle that has been integrated out. As explained in~\cite{Baumann:2011su}, this lower scale appears as a result of the $\rho \gg m$ hierarchy, which creates the window with a nonlinear dispersion for $c_s m < \omega < \rho$. The relevance of the new scale $c_sm$ can be seen, for instance, by the presence of a pole at negative frequencies in the scattering amplitude (see Fig.~\ref{contours2}). \vskip 4pt
To match the parameters of the model to the EFT parameters in Section~\ref{sec:application}, we note that
\begin{align}
\Mp^2 |\dot H| &= \frac{1}{2} |\dot \phi_0|^2\ , \qquad f_\pi^4 = \frac{m}{\rho}|\dot \phi_0|^2\ ,   \qquad \Lambda^4 =  \frac{m^5}{\rho^5}|\dot \phi_0|^2\ .\label{scal2}
\end{align}
We now compute $\pi \pi  \hskip -1pt  \to \hskip -1pt \pi \pi$  scattering in the $\pi \sigma$-model and show how it fits into the analysis of the previous sections.
 
\subsection*{$\boldsymbol{\mu=0}$}

Let us first consider the special case $\mu=0$. When $s \ll c_s^2 m^2$, the $2\to 2$ scattering amplitude for the gapless mode of the system should match the results of \S\ref{ssec:positivity}, after using \eqref{cseq} and \eqref{scal2}.  The calculation of the amplitude is technically straightforward and is performed in Appendix~\ref{sec:example}.  The result in the forward limit is 
 \beq\label{equ:egamp1}
\A \,=\, - \frac{Z^4(\omega) }{M^2}  \left\{  (\omega^2+k^2)^2  \left[ \frac{1}{4\omega^2-m^2 -  \rho^2 }  -  \frac{1}{4k^2 + m^2 } \right] - (\omega^2-k^2)^2  \frac{1}{m^2} \right\}\ , 
\eeq
where $Z(\omega)$ is the relative normalization between ${\bar \pi}$ and the scattering state of the gapless mode, and is given in~(\ref{equ:Z}).  In the low-energy limit, $k \ll m$, we have $Z(\omega \to 0) \to c_s$, and it is easy to check that the result in (\ref{equ:egamp1}) matches\footnote{To directly compare the results one must account for the rescaling $\tilde x^i = x^i/c_s$ that we used previously.} the scattering amplitude computed in the EFT after expanding in $k/m$.\vskip 4pt

For $\mu = 0$, we have $M_{n > 2}^4 = 0$ and the amplitude trivially satisfies the positivity constraint.  Nevertheless, the analytic structure and the validity of the sum rule~(\ref{equ:sumrule4}) arise quite non-trivially.  Equation~(\ref{equ:egamp1}) has two poles, one at $s = m^2 + \rho^2 \simeq \rho^2$ and another at $s= -\frac{3}{4} c_s^2 m^2 $ (or $k^2 = - \frac{1}{4}m^2$).  These two poles are related by crossing symmetry, but the pole on the negative axis is shifted relative to the location of the new physical state on the positive axis.  In the limit $c_s \ll 1$, the pole at $s= -\frac{3}{4} c_s^2 m^2 $ dominates the right-hand side of (\ref{equ:sumrule4}).
In fact, it is the only contribution at leading order in $c_s$ (see Appendix~\ref{sec:example}).

\subsection*{$\boldsymbol{\mu \neq 0}$}

For $\mu \ne 0$, we will generate non-zero $M_3^4$ and $M_4^4$ after integrating out $\sigma$.  For sufficiently large $\mu$, we expect the low-energy contributions to $M_3^4$ and $M_4^4$ to dominate over $M_2^4$.  In the following, we will work in the same limit as at the beginning of \S\ref{ssec:positivity}, namely $|c_4| \gg |c_3| \gg c_2$.
This case is particularly interesting because not every choice of $c_3$ and $c_4$ is consistent with positivity. As a result, this case offers a non-trivial test of our bounds.

The most reliable way to determine the low-energy behavior is to compute the forward amplitude and match to the EFT at low energies. This calculation is performed in Appendix~\ref{sec:example}. At leading order in $c_s \ll 1$, the amplitude in the low-energy limit, $s \ll \Lambda^2$, becomes
\beq\label{equ:muszero}
\A_{\mu^2} \,\to\, \frac{1}{8}\frac{ \mu^2}{ m^6}\, s^2  \ ,
\eeq
which matches the energy scaling of the EFT computation, as it should. More importantly, the result in (\ref{equ:muszero}) is manifestly positive. This means that any choice of $\mu$ will produce a combination of $c_3$ and $c_4$ which is consistent with the bound in \eqref{c4c3}:
\beq
c_4 - (2c_3)^2 \,=\, \frac{1}{8}\frac{\mu^2 M^2}{m^4} > 0 \ .
\eeq
Although expected, the result is non-trivial. Naively, it might have seemed possible\footnote{Given that the potential for $\sigma$ is unstable without including a quartic interaction, one might have imagined that positivity of the amplitude is enforced through stability. Perhaps unsurprisingly, positivity is a more robust feature of perturbation theory that holds for any $\mu$.} that the cubic interaction would generate large values for $c_3$ ($\gg 1$), while keeping $c_4 = 0$. This, however, would be inconsistent with positivity. 

\vskip 4pt
We have found that the $\pi\sigma$-model always produces a value of $c_4$ that is in agreement with our consistency condition. As we discuss in more detail in Appendices~\ref{sec:analytic} and \ref{sec:example}, this is a generic feature of a large class of models; in particular, this holds for all weakly coupled theories in which the $2\to 2$ scattering of the gapless mode is dominated by the exchange of a single heavy state at low energies.

\subsection[Conjecture for $c_s = 1$]{Conjecture for $\boldsymbol{c_s = 1}$}
\label{sec:cs=1}

Single-field slow-roll inflation famously leads to $c_s =1$ and produces little non-Gaussianity~\cite{Maldacena:2002vr}.  
In fact, in the flat space and decoupling limits that we have been discussing, the Lagrangian for slow-roll inflation becomes that of a free field, ${\cal L} = -\tfrac{1}{2} (\partial \phi )^2$.  This theory trivially saturates our positivity constraints because $M_{n\geq2}^4 = 0$ and $\A(s) = 0$.
However, while slow-roll inflation is consistent with our bound, it is difficult to find an explicit example of a UV-complete theory with $c_s =1$, but $c_3, c_4 \neq 0$. (For example, in DBI inflation we have $c_{n\geq 2} \to 0$ when $c_s \to 1$.)
In this section, we will provide suggestive evidence for the conjecture that theories with $c_s=1$ are always UV-completed by slow-roll inflation, without higher-order Goldstone self-interactions.
 If proven, such a result would allow us to directly link constraints on $c_s$ to the unique mechanism for inflation.   
 
\vskip 4pt
First, we will show that the positivity bound  from the previous section, $c_4 + 1>  (2c_3 + 1)^2$ (for $c_s=1$), is weaker than the constraint that derives from imposing subluminal speed of propagation in non-trivial backgrounds.
For this purpose, we return to the Goldstone Lagrangian in the form
\begin{equation}
\frac{{\cal L}}{f_\pi^4}= - \frac{1}{2} (\partial\pi)^2+\sum_{n=3}^\infty \frac{c_n}{n!}\left[-2\dot{\pi}+(\partial\pi)^2\right]^n\ ,
\end{equation}
where we have set $c_2 = 0$ since we are concerned with the $c_s = 1$ limit. 
A trivial solution to the linearized equations for motion is $\pi=\alpha_\mu {x}^\mu + \beta$.  For timelike ${x}^\mu$, we can choose $\alpha_\mu=(\alpha,0,0,0)$ and $\beta=0$. 
 At leading order in small $\alpha$, the quadratic Lagrangian for the fluctuations, $\varphi$, 
around this background (i.e.~$\pi = - \alpha t + \varphi$) is given by \begin{equation}\label{L2N}
\frac{{\cal L}_2}{f_\pi^4} = - \frac{1}{2} (\partial\varphi)^2+ 4 \alpha c_3 \hskip 1pt \dot \varphi^2 + {\cal O}(\alpha^2) \ ,
\end{equation}
where we have dropped total derivative terms. Around the new background, the speed of propagation is 
\beq
c_{s, \varphi}^2 =1 + 8 \alpha c_3 + {\cal O}(\alpha^2) \ . 
\eeq
Since $\alpha$ can have either sign, we require $c_3 = 0$ to avoid superluminal speed. 
Going to next order in $\alpha$, we find 
\beq
c_{s, \varphi}^2 =1 - 4 \alpha^2 c_4 + {\cal O}(\alpha^3) \ ,
\eeq
and superluminality is avoided iff $c_4 \ge 0$. It may be surprising that in this limit the constraint from subluminality ($c_3=0$, $c_4 \ge 0$) is stronger than that from positivity ($c_4 + 1>  (2c_3 + 1)^2$).  However, a similar observation was made in~\cite{Nicolis:2009qm}. In a relativistic EFT,  it was observed that positivity of forward scattering gave qualitatively different bounds from requiring subluminal propagation around non-trivial backgrounds, and stronger results could be derived from sum rules involving non-forward scattering amplitudes.  This suggests that a stronger bound may arise for fixed-angle scattering. 

\vskip 4pt
Inspection of the full amplitude computed in (\ref{equ:cubeamp}) and~(\ref{equ:quadamp}), shows that the only term with  angular dependence is the one proportional to $\beta_3 \equiv \frac{1}{8} (1- c_s^2)$.  
This d-wave contribution can be isolated for instance by decomposing the amplitude in partial waves, cf.~(\ref{equ:Mst}), such that $a_2(s) \propto (1-c_s^2)s^2$.
One may then hope to derive a sum rule for the d-wave amplitude (and hence the value of $c_s$):
\beq
\frac{1-c_s^2}{c_s^4}\, \stackrel{?}{=}\, \int \d s\, f(s) \ , \label{sec:cssum}
\eeq
where the function $f(s)$ would be related to the partial wave amplitudes. Isolating partial waves via non-forward dispersion relations is common in relativistic theories (e.g.~\cite{Roy:1971tc}), so it seems feasible to derive a similar expression in the non-relativistic regime.  Positivity of the sum rule \eqref{sec:cssum}, would simply correspond to subluminality of the speed of propagation at low energies, as is expected for all consistent (and Lorentz-invariant) UV theories.  At the same time, provided the right-hand-side of \eqref{sec:cssum} is positive, the vanishing of the left-hand-side for $c_s=1$  would imply that $f(s)$ must vanish.\footnote{Ideally, the function $f(s)$ would be linked to the imaginary part of the partial wave amplitude which, due to unitarity and the optical theorem, carries information about the scattering and production of intermediate states  in an interacting theory. A vanishing imaginary part would correspond to a free theory.}  This would be true (almost by definition) for a free theory, which would then constrain all interactions of the EFT to vanish.  Hence, it seems likely that a sum rule which isolates $\beta_3 = 0$ (or $c_s=1$) would ultimately force $c_{n>2} =0$.
Unfortunately, writing a sum rule for the partial waves introduces new challenges that are not present for the full amplitude at forward scattering.  First of all, the analytic properties of the scattering amplitude are less understood for non-relativistic scattering at fixed angle (or fixed transfer momentum).  Furthermore, going from the amplitude to the partial waves requires an integration over angles, which in many cases alters the (non-)analytic behavior.  Some of these shortcomings may be circumvented in the relativistic context, mostly because of the extensive use of ($s$,\hskip 2pt$t$,\hskip 2pt$u$) crossing symmetry~\cite{Roy:1971tc}, which is not available in non-relativistic theories.  

An alternative is to 
adapt
 the derivation of the Kramers-Kronig relation for the refraction index, $n(\omega)\equiv c_s^{-1}(\omega)$, to our case. 
 If $n(\omega)$ is analytic in the upper-half plane (as it is for light in a medium), and it satisfies the limit $n(\omega \gg \Lambda) \to 1$, then 
 the equivalent of the Kramers-Kronig dispersion relation holds 
\beq
{\rm Re}[n(0)] - 1\ \stackrel{?}{=}\  \int_0^{\infty} \frac{{\rm Im}[n(\omega)]}{\omega}\ . \label{equ:KK}
\eeq
We notice that (\ref{equ:KK}) is qualitatively similar to~(\ref{sec:cssum}).
In particular, if $c_s=1$ at low energies (i.e.~${\rm Re}[n(0)] \to 1$), then the dispersive term on the right-hand side again vanishes.
One of the obstacles in this derivation is establishing the off-shell frequency/momentum dependence of the Green's function.  Although causality guarantees certain properties for the Green's function,  these are not  translated as easily into the analytic behavior of $n(\omega)$ as in the electromagnetic case. 
While we do not think that the problems described above are insurmountable, they make the status of non-relativistic sum rules for partial wave amplitudes, or the refraction index, somewhat uncertain. We will return to these issues in future work.

\section{Conclusions}
\label{sec:conclusion}

Observations of the CMB anisotropies can be traced back to the moment of horizon crossing during inflation.
These observations therefore probe energies of order the inflationary Hubble scale.  One of the key challenges in cosmology is to relate these measurements to the microphysics of inflation, which is separated from the Hubble scale by a sizable energy gap, e.g.~$(H/f_\pi)^2 \simeq 10^{-4}$. 
In this paper, we used causality (and unitarity) to link cosmological observables, and the related coefficients in the IR theory, to the unknown UV~dynamics of inflation.

\begin{figure}[h!]
\centering
\includegraphics[scale=0.6]{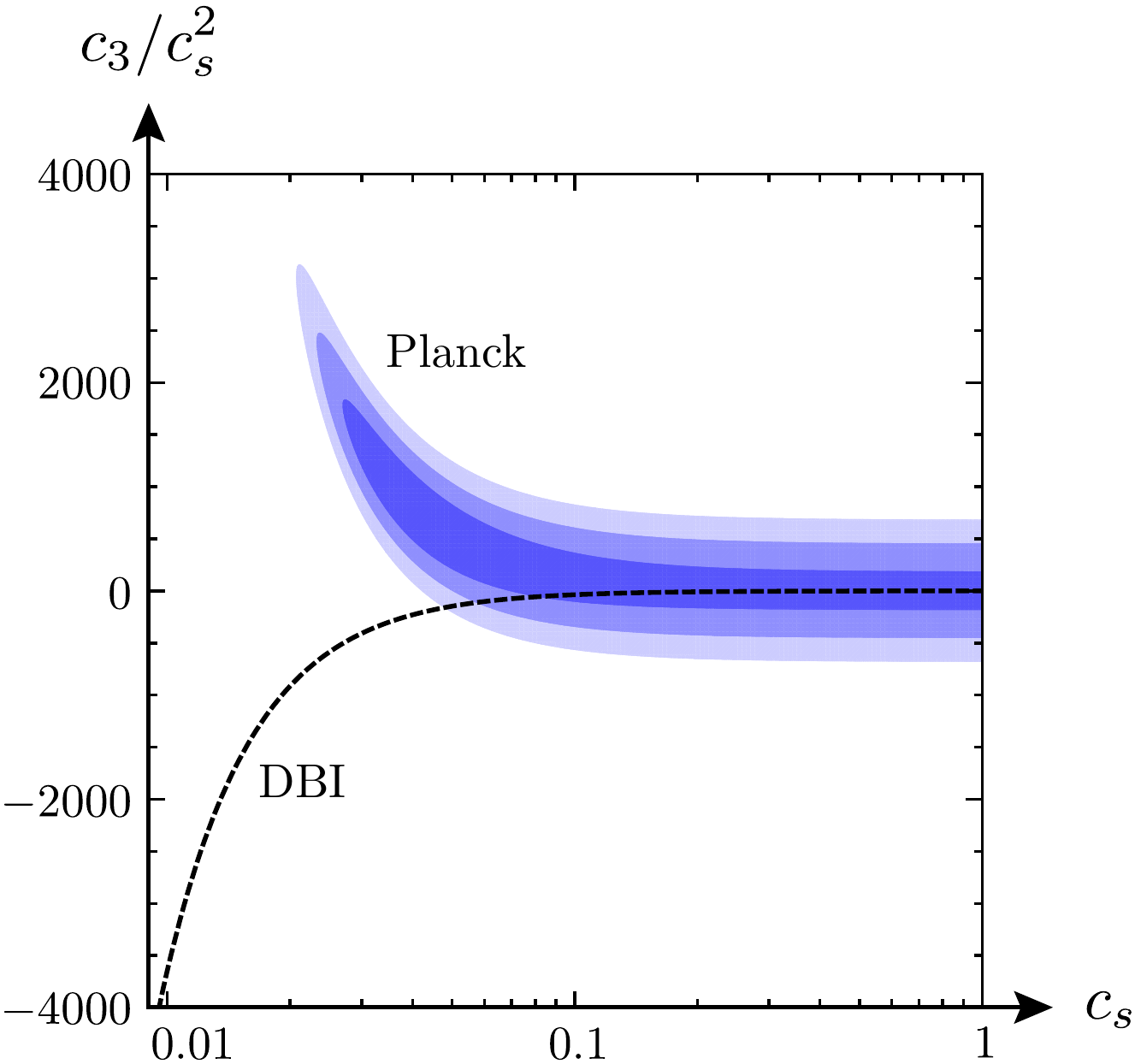}
\caption{Observational constraints on the EFT parameters $c_s$ and $c_3$~\cite{PlanckNG}. }
\label{fig:constraints}
\end{figure}

\vskip 4pt
The information that can be extracted from the low-energy measurements is limited. 
The two-point function of temperature fluctuations measures the amplitude ($A_s$) and the scale-dependence~($n_s$) of the primordial scalar perturbations, and puts a bound on the amplitude of tensor modes, often quoted as the tensor/scalar ratio ($r$).
Higher-order correlations, in principle, measure (or constrain) additional parameters.
For single-field inflation, these parameters include the sound speed~($c_s$), as well as a cubic coupling ($c_3$) and a quartic coupling ($c_4$). The latest constraints on the parameters $c_s$ and $c_3$ from the CMB bispectrum~\cite{PlanckNG} are shown in fig.~\ref{fig:constraints}.  The first constraint on the parameter $c_4$ has recently been derived from measurements of the CMB trispectrum~\cite{PlanckNG} (see also~\cite{Smith:2015uia, Fergusson:2010gn, Regan:2013jua}) 
\beq
- 8.3 \times 10^7  \,<\, c_4/c_s^4 \,<\, 7.4 \times 10^7 \quad (95\%\hskip 1pt\rm CL).   \label{equ:c4exp}
\eeq
Let us note that this limit assumes $c_3 = 0$, and a dedicated analysis of the CMB bispectrum and trispectrum for general $c_s$, $c_3$ and $c_4$ is still lacking.  However, we already see that much of the parameter space remains to be observationally explored. The theoretical bounds that we discussed in this paper are therefore very relevant.

\vskip 4pt
We showed that analyticity of the $2 \to 2$ scattering amplitude for the Goldstone boson implies a sum rule that relates a combination of the parameters ($c_s$, $c_3$, $c_4$) to an integral over the high-energy spectrum of the scattering amplitude: cf.~\eqref{equ:sumrule4}.  Hence, the EFT parameters are connected to specific features of scattering processes in the UV completion of inflation. 
Assuming positivity of the sum rule, we then derived a new consistency condition which bounds the size of the  four-point function in terms of the square of the three-point function for equilateral configurations. This consistency condition restricts the size and the sign of the quartic coupling~$c_4$.  
On purely theoretical grounds, we have thus ruled out about half of the parameter space allowed by~(\ref{equ:c4exp}), for all known UV completions of the EFT of inflation (and extensions thereof). While we have not been able to construct an explicit example in which our bound is violated, we have isolated the necessary ingredients. We have also argued that our consistency condition is a generic feature in weakly coupled theories. Hence, finding large negative values of $c_4$ would point towards less conventional (plausibly strongly coupled) theories of inflation, or more radically to violations of basic properties of scattering amplitudes (e.g.~\cite{smatrix}).

\vskip 4pt
We consider the present work to be only a first and modest step towards a more complete understanding of the IR/UV connections between cosmological observations and the underlying physics of inflation.  Many future directions suggest themselves. For instance, we may hope to find sum rules for individual parameters of the EFT, rather than just for a special combination of several of them.  This may be possible by extending our analysis to non-forward scattering, or through generalized Kramers-Kronig relations for the Green's functions. We have speculated that such an analysis would allow us to derive a sum rule for $c_s$, the speed of propagation of the Goldstone mode. In this case, positivity would correspond to the expected subluminality condition: $c_s < 1$. On the other hand, in the limit $c_s \to 1$, the sum rule would constrain the total amplitude to vanish. This has lead us to conjecture that theories with $c_s=1$ can only be UV completed by slow-roll inflation. While, so far, we have only given suggestive evidence for this intriguing conjecture, we hope to provide a positive answer to this question in the near future.
\vskip 1cm
\begin{center}
{\bf Acknowledgements}
\end{center}
We thank Valentin Assassi,  Brando Bellazzini, Cliff Burgess, Liam Fitzpatrick, Garrett Goon, Kurt Hinterbichler, Takeshi Kobayashi, Juan Maldacena, Joel Meyers, Guilherme Pimentel, Yi Wang and Matias Zaldarriaga for discussions and comments. D.B.~thanks the theory groups at CERN (Geneva), IAP (Paris), and IoP (Amsterdam) for their hospitality.  R.A.P.~thanks the theory groups at ASC~(Munich), Boston University, DESY (Hamburg) and KITP (Santa Barbara) for their hospitality while this work was being completed. D.B. and H.L.~acknowledge support from a Starting Grant of the European Research Council (ERC STG grant 279617). H.L.~is supported by the Cambridge Overseas Trust, the Lord Rutherford Memorial Research Fellowship, the Sims Empire Scholarship, and the William Georgetti Scholarship.
D.G. is supported by a NSERC Discovery grant. R.A.P. is supported by the Simons Foundation and the S\~ao Paulo Research Foundation (FAPESP) under grants 2014/25212-3 and 2014/10748-5.

\newpage
\appendix
\section{Analyticity of the Scattering Amplitude}
\label{sec:analytic}

In this appendix, we discuss further the analytic properties of $2\to2$ scattering of identical $\pi$-particles. 

\vskip 4pt
Without loss of generality, the non-forward scattering amplitude $\M$ may be written as a function of the following variables:
\begin{align}
\omega_{12}   &\equiv \omega_{1} +\omega_2\ ,  \qquad \vec k_{12}  \equiv \vec k_1 + \vec k_2\ , \nonumber \\
\omega_{13}   &\equiv \omega_{1} -\omega_3\ ,  \qquad \vec k_{13}  \equiv \vec k_1 - \vec k_3\ , \nonumber \\
\omega_{14} &\equiv \omega_{1} - \omega_4\ , \qquad \vec k_{14} \equiv\vec k_1 - \vec k_4\ . \label{w12}\
\end{align}
In the UV, i.e.~for $\omega_a \gg \rho$, we expect the amplitude to become a function of the standard Mandelstam variables ($s$,\hskip 2pt$t$,\hskip 2pt$u$). Moreover, in the IR, some of the contributions to $\M$ may simplify to expressions in terms of the re-defined Mandelstam variables ($\tilde s,\hskip 1pt\tilde t,\hskip 1pt\tilde u$) associated with the re-scaled momenta, $\tilde p_a \equiv (\omega_a, c_s(\omega_a)\hskip 1pt \vec{k}_a)$. These contributions come from the terms in the effective action that mimic relativistic interactions after the rescaling of the spatial coordinates, e.g.~$(\tilde\partial\pi)^4$. 
The general expression for $\M$ in the non-relativistic regime may (and will) 
contain additional Lorentz symmetry breaking combinations. 

\vskip 4pt
At low energies, the scattering amplitude computed in the EFT description must, of course, match the one computed in the full theory. Analyzing this matching in general may be cumbersome. However, for forward scattering in the center-of-mass frame ($\vec k_{ab}=0$, $\omega_{13}=\omega_{14}=0, \omega_{12} \equiv 2\omega$) some simplifications occur. In particular, the amplitude $\A_{\rm cm}$ can be expressed in terms of the square of the center-of-mass energy, $\omega_{12}^2 = 4\omega^2$, which in this frame is equal to both $s$ and~$\tilde s$.  
For notational simplicity, we will write the forward scattering amplitude in the center-of-mass frame as $\A_{\rm cm}(s) \equiv \A_{\rm cm}(4\omega^2)$. In the main text, we dropped the subscript `cm', but here we keep it explicit  in order to highlight expressions which are only valid in a fixed frame. The distinction becomes important when studying the implications of crossing symmetry, since these are better described in a frame-independent manner and dropping the `cm' subscript could lead to confusion.

\vskip 4pt 
The standard properties of the relativistic formalism (cf.~\S\ref{sec:rel}) apply to the full amplitude in the UV. This means that any singularities in $\A_{\rm cm}(s)$ off the real axis, if present, would have to come from the non-relativistic IR behavior of the amplitude. 
On the one hand, for positive real~$s$, the argument that restricts the non-analytic behavior to a minimum (to be consistent with unitarity and the optical theorem) remains unchanged. Moreover, for $s < -\rho^2$, 
crossing symmetry relates the amplitudes in the $s$- and  $u$-channels, where similar considerations apply. On the other hand, for $-\rho^2  < s < 0$, we will demonstrate that crossing symmetry does not simply relate $\A_{\rm cm}(s)$ to $\A_{\rm cm}(-s)$, as in the relativistic case. However, except for some rather peculiar behavior, which we will discuss later, singularities for unphysical values of $s$ will be associated with physical poles and/or branch cuts for physical values of $s$ (albeit not directly symmetric points). 
We therefore do expect the Mandelstam hypothesis of maximal analyticity to hold, and any non-analytic behavior to be restricted to the real $s$-axis.  
It remains to be analyzed whether these singularities along the negative real axis, especially in the region $ -\rho^2 < s < 0$, could jeopardize positivity of the sum rule discussed in \S\ref{ssec:NR}. As we shall see, crossing symmetry still plays a major role in determining the location of the non-analytic behavior.

 \vskip 4pt
In quantum field theory, crossing symmetry follows from the properties of the Green's functions and the LSZ reduction formula \cite{weinberg2005theV1}. Put simply, field operators may create an incoming particle or an outgoing anti-particle out of the vacuum. For a relativistic theory with identical scalar particles, it is easy to use this property to connect regions of the scattering amplitude when $(s,t,u)$ are exchanged. For non-relativistic theories, the LSZ formula still applies at low energies, but the relation between the different channels becomes more subtle. 
In particular, for energies near the cutoff $\Lambda$ and below the UV scale $\rho$, extra poles or cuts may develop. The computation in terms of field operators implies that the crossing symmetry between the $s$- and $u$-channels relates the scattering amplitude under the exchange $\omega_{2} \leftrightarrow -\omega_{4}$ and $\vec{k}_2 \leftrightarrow - \vec{k}_4$.
At forward scattering, this transformation implies
\beq
 \tilde s \equiv \omega_{12}^2 - c_s^2 {\vec k_{12}}^{\hskip 1pt 2} = 4 \omega^2 \ \leftrightarrow\ \tilde u \equiv \omega_{14}^2 - c_s^2 {\vec k_{14}}^{\hskip 1pt 2} = -4 \omega^2 = - \tilde s\ ,
\eeq
where we have evaluated the expressions on-shell.  The part of the amplitude that is only a function of $\tilde s$ (in a generic frame) is therefore an even function of $\tilde s$.
However, in principle the scattering amplitude~$\M$ also has contributions that do not transform as easily. For instance, $\omega_{12} \leftrightarrow \omega_{14}$ under the crossing symmetry, but $\omega_{14}$ vanishes in the center-of-mass frame, while 
$\tilde s = \omega_{12}^2$ prior to the crossing transformation. Terms that vanish in the center-of-mass frame, e.g.~those proportional to $\omega_{14}$, play a vital role in making crossing symmetry manifest. For this reason, it is useful to distinguish functional dependence on $\tilde s$ from explicit functions of~$\omega_{12}$.

\vskip 4pt
To illustrate these considerations, let us study the exchange of a heavy state 
 in the $s$-channel, away from forward scattering and in a generic frame. Using standard `polology' arguments \cite{weinberg2005theV1}, we expect the amplitude to take the following form  
\beq
\M_s\,\supset\, \frac{\Z(\omega_{ab},\vec k_{ab}\cdot \vec k_{cd})}{\omega_{12}^2 - c_r^2\hskip 1pt \vec k_{12}^{\hskip 1pt 2} - M^2+ i \epsilon} \ , \label{equ:A10}
\eeq
where $M$ is the energy of the intermediate state, $c_r \equiv c_r(\omega_{12})$ is its speed of propagation, and $\Z$ is some unknown function of the quantities defined in \eqref{w12}. This expression must be symmetric with respect to permutations of the momenta that leave the $s$-channel fixed: i.e.~$\{1 \leftrightarrow 2 \}$ and $\{3 \leftrightarrow 4\}$.  
It is useful to write the amplitude in terms of variables that make this invariance manifest (after using the on-shell conditions), namely\footnote{To avoid a proliferation of different names we abuse notation and denote both functions in (\ref{A4}) by $\Z(\cdots)$.}
\beq
  \Z(\omega_{ab},\vec k_{ab}\cdot \vec k_{cd}) \equiv \Z( \tilde s, \hskip 1pt  \omega_{12}^2, \hskip 1pt  \omega_{13}^2+\omega_{14}^2, \hskip 1pt  \vec k_{13}\cdot  \vec k_{14}, \hskip 1pt \omega_{13}\hskip 1pt\omega_{14}) \ , \label{A4}
\eeq
where we have chosen to express $\vec k^{\, 2}_{12}$ in terms of $\tilde s$ and $\omega_{12}$.  We can then use crossing symmetry to determine the location of the pole in the $u$-channel, which we denote by~$\M_u$. Putting both contributions together, we find 
\begin{align}
\M  &\,=\, \M_s+ \M_u \nonumber \\[4pt]
&\,\supset\, \frac{\Z( \tilde s,  \hskip 1pt  \omega_{12}^2,  \hskip 1pt   \omega_{13}^2+\omega_{14}^2,  \hskip 1pt  \vec k_{13}\cdot  \vec k_{14},  \hskip 1pt  \omega_{13} \hskip 1pt \omega_{14}) }{\omega_{12}^2 - c_r^2 \hskip 1pt \vec k_{12}^{\hskip 1pt 2} - M^2 + i \epsilon} \,+\,   \frac{\Z( \tilde u,  \hskip 1pt  \omega_{14}^2,  \hskip 1pt   \omega_{13}^2+\omega_{12}^2,  \hskip 1pt  \vec k_{13}\cdot  \vec k_{12},  \hskip 1pt \omega_{13} \hskip 1pt \omega_{12})}{\omega_{14}^2 - c_r^2 \hskip 1pt \vec k_{14}^{\hskip 1pt 2} - M^2 + i \epsilon} \ .
\end{align}
Taking the forward limit, ${\vec k}_{13} \to 0$ --- but still in a generic frame --- we get  
\beq
\A \,\supset\, \frac{\Z( \tilde s,  \hskip 1pt  \omega^2_{12},  \hskip 1pt  \omega_{14}^2) }{\omega_{12}^2 - c_r^2 \hskip 1pt \vec k_{12}^{\hskip 1pt 2} - M^2+ i \epsilon} +   \frac{\Z(-\tilde s,  \hskip 1pt \omega^2_{14},  \hskip 1pt  \omega_{12}^2) }{\omega_{14}^2 - c_r^2 \hskip 1pt \vec k_{14}^{\hskip 1pt 2} - M^2 + i \epsilon} \ ,
\label{eqnA}
\eeq
where $\Z(x,y,z)\equiv \Z(x,y,z,0,0)$. At high energies, $\omega \gg \rho$, 
Lorentz invariance is restored and we expect crossing symmetry to act in the familiar way.
To see this, we note that $c_s(\omega), c_{r}(\omega) \to 1$ and $\tilde s \to s$ in the UV. Moreover, the amplitude will be dominated by a relativistically invariant function,  
\beq
\Z(\tilde s, \ldots) \, \xrightarrow{\ \omega \gg \rho \ } \, \Z(s, \ldots) = \Z^{\rm UV}(s)\big(1+ {\cal O}(\rho/\omega)\big)\, ,
\eeq
where $\Z^{\rm UV}(s) = \Z^{\rm UV}(-s)$, as required by crossing symmetry when the theory becomes relativistic. As expected, the expression in \eqref{eqnA} therefore becomes symmetric under $ s \to - s$ (and $\epsilon \to -\epsilon$). This is also manifest in the center-of-mass frame, where we have   \beq
\A_{\rm cm}(s) \,\supset\, \frac{\Z^{\rm UV}_{\rm cm}(s)}{s  - M^2 + i \epsilon} +  \frac{\Z^{\rm UV}_{\rm cm}(-s) }{-s - M^2 + i \epsilon} \ , ~~~~~\text{for $s\gg \rho^2$\,.} \label{equ:A14n}
\eeq

At low energies, on the other hand, crossing symmetry does not guaranteed that $\A_{\rm cm}(s)$ is an even function of $s$. Instead, we have
\beq
\A_{\rm cm}(s) \,\supset\, \frac{\Z_{\rm cm}(s , s, 0) }{s  - M^2 + i \epsilon} +   \frac{\Z_{\rm cm}(-s , 0, s) }{-c_r^2 c_s^{-2}\hskip 1pt s - M^2 + i \epsilon} \ , ~~~~~\text{for $s < \rho^2$\,,} \label{equ:A14}
\eeq
and the two terms are not necessarily related by reflection.  
First of all, when $c_r \neq c_s$, the location and residue of the pole on the negative axis is not the symmetric counterpart of the one on the positive axis. 
(This is seen explicitly in the perturbative example discussed in \S\ref{sec:WeakExample} and Appendix~\ref{sec:example}; cf.~fig.~\ref{contours2}.)
Furthermore, while the optical theorem forces the residue of the $s$-channel pole in \eqref{equ:A14} to be positive, this does not imply positivity of the residue of the $u$-channel pole. Unitarity alone is not sufficient to guarantee positivity because the function $\Z_{\rm cm}(x,y,z)$ is evaluated for different arguments in the $s$- and $u$-channels.\vskip 4pt  We may then worry that the residue from the negative $s$-axis may be negative and dominate over the positive contribution from the $s$-channel. Fortunately, in many circumstances we find that $\Z$ is 
invariant under permutations of~$\omega_{a}$, such that 
\beq 
\Z \to \Z(\tilde s, \omega_{1}\omega_2\omega_3\omega_4)\ .
\eeq 
For example, this property arises when time derivatives act on the external legs. In the center-of-mass frame, this means that\footnote{In a perturbative setting, attempts to put an unequal number of time derivatives on each leg fail to produce singularities from the $u$-channel in the center-of-mass frame.  After summing over permutations, the $u$-channel amplitude becomes a function of $\omega_{14} $, which vanishes when $\omega_1 =\omega_4=\omega$.}
\beq 
\Z(\tilde s, \omega_{1}\omega_2\omega_3\omega_4) \,\to\, \Z_{\rm cm} (s, \omega^4) = \Z_{\rm cm} (s, s^2)\ ,
\eeq 
which extends the original form of the crossing symmetry to all energies. The residues on the positive and negative $s$-axes are therefore related, and both constrained to be positive by the optical theorem. This is also manifestly true in the example of \S\ref{sec:WeakExample} and a large class of weakly coupled extensions.
\vskip 4pt
The above reasoning takes into account poles and branch cuts that originate in the $s$-channel. There is, however, a final subtlety to be discussed. Since Lorentz invariance is broken, interactions can in principle have an unequal number of time and space derivatives. For example, a quartic interaction with three time derivatives and two spatial derivatives can be consistent with the symmetries of the EFT. This could in principle produce contributions in the IR of the form
\beq
\A \propto \omega_{12}^{5} = (\omega_{12}^2)^{5/2} \to \A_{\rm cm} \propto s^{5/2} \ .
\eeq
To be consistent with unitarity,  
we must choose these cuts to run along the negative axis.
 Notice, however, that in a generic frame crossing symmetry maps $\omega_{12} \to \omega_{14} \to 0$. Therefore these type of singularities do not have an $s$-channel counterpart. While potentially dangerous, these terms are always subdominant in perturbation theory, since they involve higher powers of $s$. They must be absent also in the UV theory, which is dominated by relativistic interactions. Hence, at lowest order in $s$, these rather peculiar singularities do not present a problem for our positivity argument.

\section{Positivity in the $\boldsymbol{\pi\sigma}$-model}
\label{sec:example}

In this appendix, we present details of the analysis of the weakly coupled example of \S\ref{sec:WeakExample}. Specifically, we will show how the sum rule (\ref{equ:sumrule4}) is realized in this particular example and demonstrate explicitly that it satisfies our positivity bound.

\vskip 4pt
For convenience, we recall the Lagrangian for the $\pi\sigma$-model:
\begin{align}
{\cal L} \,=\, - \frac{1}{2}(\partial \bar \pi)^2 -\frac{1}{2} (\partial \sigma)^2    - \rho\hskip 1pt \sigma \dot{\bar \pi} - \frac{\sigma (\partial \bar \pi)^2}{2M} - \frac{1}{2}m^2 \sigma^2 - \frac{1}{3!} \mu \sigma^3\ .
\end{align}
In the flat space limit, the linearized equations of motion are given by
\begin{align}
(\omega^2-k^2)\bar\pi +i\rho\omega\sigma &=0\ ,\\
(\omega^2-k^2-m^2)\sigma - i\rho \omega\bar\pi &= 0\ ,
\end{align} 
so that the propagator for ${\boldsymbol \phi} \equiv  (  \bar \pi \hskip 4pt \sigma)$ is 
\beq
 \langle T({\boldsymbol \phi}_p {\boldsymbol \phi}_{-p}^T) \rangle = \frac{i}{(\omega^2-k^2)(\omega^2-k^2-m^2) -  \omega^2 \rho^2 + i \epsilon} \, \left( \begin{array}{cc} \omega^2-k^2 -m^2 & - i \rho \omega \\ i \rho \omega & \omega^2-k^2 \end{array} \right)\ .\label{equ:prop}
\eeq
The poles of the propagator are associated with the non-trivial solutions for $\bar\pi$ and $\sigma$, which satisfy 
\begin{equation}
\omega^2_\pm = k^2 + \tfrac{1}{2}(\rho^2+m^2) \pm \sqrt{\rho^2 k^2+\tfrac{1}{4}(\rho^2+m^2)^2}\ .\label{equ:disprel}
\end{equation}
The mixing of $\bar \pi$ and $\sigma$ presents an additional complication because at low energies neither $\bar\pi$ nor $\sigma$ creates an energy eigenstate. To correct for this, we will compute the $S$-matrix elements using the LSZ formula~\cite{Peskin}:  
\beq
S \,=\, \left(\,\prod_{a=1}^4\, \lim_{\omega_a \to E_a } \frac{\omega_a^2 - E_a^2}{Z(\omega_a)} \right) \langle T  (\bar\pi_{p_1}\bar\pi_{p_2} \bar\pi_{p_3}\bar \pi_{p_4}) \rangle \ ,
\eeq
where $E_a$ is the energy of the gapless state. The function $Z(\omega)$ is the relative normalization between $\bar\pi$ and the canonically-normalized  energy eigenmode,\footnote{Let us emphasize this is just a choice. In principle, the scattering amplitude can be computed using any interpolating field with a non-vanishing overlap with the asymptotic eigenstates, see e.g.~\cite{weinberg2005theV2}. In our case it turns out to be convenient to work directly with $\bar\pi$ rather than diagonalizing the propagator in~\eqref{equ:prop}.}  
\beq
\langle T(\bar  \pi_{p}  \bar \pi_{-p}) \rangle = \frac{Z^2(\omega)}{\omega^2 - E^2( k) }\ ,
\eeq
which in this particular case is given by
\beq\label{equ:Z}
Z( \omega_a = E_a( k_a) ) = \left(\, \frac{m^2 -\rho^2 + \sqrt{4 k^2  \rho^2 + (m^2 +\rho^2)^2 }}{2 \sqrt{4 k^2  \rho^2 +  (m^2 +\rho^2)^2 }}\, \right)^{1/2} \ .
\eeq

\subsection*{$\boldsymbol{\mu=0}$}

We first consider the special case $\mu=0$.  The forward scattering amplitude gets contributions from exchange diagrams that include all the matrix elements in the propagator~(\ref{equ:prop}).  
There are three classes of these diagrams: $\sigma$-exchange, $\pi$-exchange and $\pi\sigma$-exchange.  At low energies, and for $c_s \ll 1$, the $\sigma$-exchange contribution dominates the amplitude. In the forward limit, we then find the following amplitude in the center-of-mass frame
\beq
\A \,=\, - \frac{Z^4(k^2)}{M^2} \left\{  (\omega^2+k^2)^2  \left[ \frac{1}{4\omega^2-m^2 -  \rho^2 }  -  \frac{1}{4k^2 + m^2 } \right] - (\omega^2- k^2)^2  \frac{1}{m^2} \right\}\ , \label{AMP}
\eeq
where the last term is from the $t$-channel exchange. In the limit $\omega \to 0$, 
this amplitude indeed matches the result of the EFT computation. We see that the amplitude has poles at $4\omega^2 = m^2 + \rho^2 \simeq \rho^2$ and $4k^2 = - m^2$ (or $4\omega^2 \simeq - \frac{3}{4} c_s^4\rho^2$); cf.~fig.~\ref{contours2}.
For small $c_s$, the pole on the negative axis is located much closer to the origin than that on the positive axis.
 
 \vskip 4pt
We wish to see how the sum rule (\ref{equ:sumrule4}) works for the amplitude (\ref{AMP}).  It is easy to see that the residue of the pole on the negative axis dominates: the pole on the positive axis is suppressed by a factor of $c_s^2$, while the relativistic regime, $M > \omega \gg \rho$, only contributes $ \ln(M/\rho)/M^4$.
Using that the imaginary part associated with the pole on the negative axis is 
\beq
{\rm Im}[\A(s<0)] =  \frac{Z^4(k^2)}{M^2}\, (\omega^2+k^2)^2 \,  \pi \, \delta(- 4 k^2 - m^2) \ , \label{equ:ImAneg}
\eeq   
the sum rule can be written as
\beq
\frac{1}{2}\A^{\prime \prime}(s \to 0) = \frac{1}{\pi} \int_{-\infty}^{0} \frac{\d s}{s^3}\, {\rm Im}[\A(s)] =   \int_{0}^{\infty} \d q\, \frac{s'(q)}{s^3(q)} \frac{Z^4(q)}{M^2}  \frac{q^2}{16}\,  \delta(q-m^2)\ , \label{equ:10}
\eeq
where $q=-4 k^2$ and
\begin{align}
s(q) &\equiv - q  + 2(m^2 +\rho^2) - 2 \sqrt{-q  \rho^2 +(m^2 +\rho^2)^2 } \ , \\
s^\prime(q) &\equiv \frac{ds}{dq} = -1 + \frac{\rho^2}{\sqrt{-q  \rho^2 +(m^2 +\rho^2)^2 } }\ .
\end{align}
At leading order in $c_s = m/\rho \ll 1$, we have
 \beq
 s(q=m^2) = -\frac{3}{4} c_s^2\ , \quad s'(q=  m^2) = -\frac{3}{2} c_s^2\ , \quad Z^2(q=m^2) = \frac{3}{4} c_s^2\ .
 \eeq
Substituting this into (\ref{equ:10}), we find 
\beq\label{equ:1pole}
\frac{1}{2}\A^{\prime \prime}(s \to 0) = \frac{1}{8 m^2 M^2} 
= \frac{1}{8 |\dot{\phi}_0|^2 c_s^2}\ ,
\eeq
where we have used $m=c_s \rho$ and $\rho = |\dot{\phi}_0|/M$. 
The left-hand side of (\ref{equ:1pole}) can also be computed directly in the EFT for the canonically normalized field $\pi_c= c_s \pi / |\dot{\phi}_0|$ (after integrating out $\sigma$). In the limit $c_s \ll 1$, 
eq.~(\ref{equ:As}) becomes\footnote{In Section~\ref{sec:application}, we rescaled the coordinates by $\tilde x^i = x^i/c_s$.  This rescaling changes the normalization of the amplitude. 
We have corrected for this difference by rescaling the result of Section~\ref{sec:application} by a factor of $c_s^3$.} 
\beq
\frac{1}{2}\A^{\prime \prime}(s\to 0)  = c_s^3 \times \frac{1}{8} \frac{1}{\Lambda^4} =  \frac{1}{8|\dot{\phi}_0|^2 c_s^2 }\ ,
\eeq
where we used $\Lambda = f_\pi c_s$ and $f_\pi^4 = |\dot{\phi}_0|^2 c_s$.
We thus find exact agreement, at leading order in $c_s \ll 1$, with the single pole contribution to the dispersion relation.

\subsection*{$\boldsymbol{\mu \ne 0}$}

Finally, we compute the forward scattering amplitude for $\mu \ne 0$.
We will assume that $\mu$ is sufficiently large that we can neglect all other cubic terms.
This example generates large $c_3$ and $c_4$ in the EFT. We wish to determine whether the derived EFT parameters satisfy our positivity constraint.
A similar computation to the one above gives the ${\cal O}(\mu^2)$ contribution to the forward amplitude
\beq
\A_{\mu^2} \,=\,- \frac{ \mu^2}{Z^4(\omega)} \left(\frac{ \omega \rho}{\sqrt{4 p^2 \rho^2+ (m^2 + \rho^2)^2}}\right)^4  \left[\frac{1}{4\omega^2-m^2 -  \rho^2} -  \frac{1}{4p^2+m^2} - \frac{1}{m^2}\right] \ .
\eeq
The analytic properties of this amplitude are similar to the previous case with poles located at
$4\omega^2 = m^2 + \rho^2 \simeq \rho^2$ and $4k^2 = - m^2$.
In the limit $\omega \to 0$, we get
\beq
\A_{\mu^2} \,\to\, \frac{1}{8} \frac{\mu^2}{m^6} \, s^2\ ,
\eeq
which is manifestly positive.

\vskip 4pt
Although the analysis of this appendix was a non-trivial check of our positivity constraint, 
the underlying reason for the positivity was already anticipated in Appendix~\ref{sec:analytic}. 
Specifically, the low-energy amplitude was UV completed through the
exchange of a single heavy state. As a result, the coefficient function, $Z$, must scale as $s^2$ in order to match the low-energy scaling of the EFT.\footnote{One may also have $Z \propto s$ in such a way that the leading contributions to the $s$- and $u$-channels cancel in the limit $s\to0$, leaving $\A(s) \propto s^2$, as required.  Such a cancelation will only occur when the sign of the $u$-channel term is consistent with a positive contribution to our sum rule and therefore does not present a loophole to this argument.} 
It is clear that this scaling arises from a single
derivative acting on each external leg and therefore $Z$ is manifestly crossing symmetric. 
As a result, the residues of the $u$- and $s$-channel poles must have the same sign, and therefore the forward amplitude must be positive.  One can check that this conclusion cannot be altered by changing the form of the interactions or of the mixing term. We conclude that positivity of the sum rule is a generic feature of weakly coupled UV completions of the EFT of inflation.

\section{Low-Energy $\boldsymbol{\pi\pi \to \pi\pi}$ Scattering}
\label{sec:computation}

In this appendix, we compute the low-energy $\pi\pi \to \pi\pi$ scattering in the EFT of inflation at leading order in the derivative expansion. At tree level, we have two types of diagrams: i)~exchange diagrams involving the combination of two cubic vertices, and ii) contact diagrams involving quartic vertices.
We will treat these two scattering processes in turn.

\subsection*{Exchange diagrams}

The Lagrangian at cubic order is
\begin{align}
\tilde {\cal L}_3 =  \frac{1}{\Lambda^2} \left[ \alpha_1 \hskip 1pt \dot \pi_c^3 -\alpha_2 \hskip 1pt \dot \pi_c(\tilde \partial \pi_c)^2 \right]  \ , \label{L3B}
\end{align}
where the parameters $\alpha_i$ are defined in (\ref{equ:alphas}).
For each exchange diagram, we get factors of $\frac{1}{2}i^2$ from the two vertices, $i^3(-i)^3$ from the six momenta, and $i$ from the propagator, leading to an overall factor of $-\frac{1}{2}i$. 
The two interactions in (\ref{L3B}) lead to three different types of exchange diagrams:
\begin{itemize}
\item $\boldsymbol{\dot{\pi}^3 \times \dot{\pi}^3}$. We first consider the diagram involving two factors of the interaction $\dot \pi^3$.  The internal contraction for this diagram only involves time derivatives, which implies that only the $s$-channel is non-vanishing in the center-of-mass frame (using $\omega_{13}=\omega_{14}=0$). There are $3^2=9$ ways of choosing this internal contraction and $4\times 2=8$ diagrams for the $s$-channel; hence the symmetry factor in this case is 72. The vertices give a factor of $\alpha_1^2/\Lambda^4$, and we get
\begin{align}
i{\cal M}_{\dot{\pi}^3} = -\frac{1}{2}i\cdot 72\cdot \frac{\alpha_1^2}{\Lambda^4} \cdot \Big[\omega^2(2\omega)\Big]\,\frac{1}{s}\,\Big[\omega^2(2\omega)\Big]= -\frac{9i}{4}\, \alpha_1^2\, \frac{s^2}{\Lambda^4}\ ,
\end{align}
where the final equality holds in the center-of-mass frame.

\item $\boldsymbol{\dot{\pi}(\partial\pi)^2 \times \dot{\pi}(\partial\pi)^2}$. The computation of the diagram involving two factors of $ \dot{\pi}(\partial\pi)^2$ is slightly more involved. Now there are three possible internal contractions:
\begin{itemize}
\item $\contraction{}{\dot{\pi}}{}{\dot{\pi}}\dot{\pi}\dot{\pi}$. This internal contraction consists of time derivatives only, so only the $s$-channel survives. Since there is only one way of choosing the internal contraction, the symmetry factor is 8 and we get
\begin{equation}
i{\cal M}_{\dot{\pi}(\partial\pi)^2,a} = -\frac{1}{2}i \cdot 8 \cdot \frac{(-\alpha_2)^2}{\Lambda^4}  \Big[( \tilde p_1\cdot \tilde p_2) 2\omega\Big]\,\frac{1}{s}\,\Big[( \tilde p_3\cdot \tilde p_4) 2\omega \Big] = -i\, \alpha_2^2\,\frac{s^2}{\Lambda^4}\ .
\end{equation}
\item $\contraction{}{\dot{\pi}}{}{\partial\pi}\dot{\pi}\partial\pi$. Again, only the $s$-channel contributes, but now there are 4 possible ways of choosing the internal contraction, giving a symmetry factor of $8\times 4=32$. The amplitude is
\begin{equation}
i{\cal M}_{\dot{\pi}(\partial\pi)^2,b} = - \frac{1}{2}i \cdot 32 \cdot \frac{(-\alpha_2)^2}{\Lambda^4}  \Big[\omega \hskip 2pt \tilde p_1\cdot (\tilde p_1+\tilde p_2)\Big]\,\frac{1}{s}\,\Big[2\omega (\tilde p_3\cdot \tilde p_4)\Big] = -2i\, \alpha_2^2\, \frac{s^2}{\Lambda^4}\ .
\end{equation}
\item $\contraction{}{\partial\pi}{}{\partial\pi}\partial\pi\partial\pi$. Since there are no time derivatives appearing in the internal contraction this time, naively we would expect that both the $t$- and $u$-channels would contribute. However, it turns out that both vanish in this case too. To see this, note that the scattering amplitude in the $t$-channel contains terms such as $(\tilde p_1- \tilde p_3)\cdot(\tilde p_1+ \tilde p_3)=0$ (and similarly for the $u$-channel), giving zero amplitude.\footnote{The absence of low-energy poles is a genuine feature for all tree level exchange diagrams in the EFT of inflation, so that the forward scattering limit is well-defined in spite of $\pi$ being massless. To see this, first note that any internal contraction involving time derivative operators will vanish in the $t$-channel, and moreover those involving box operators will bring up factors of $t$, cancelling with the poles in the denominator. The remaining contractions then involve terms of the form $\partial_{\mu_1\cdots\mu_n}\pi \hskip 1pt\partial_{\nu_1\cdots\nu_m}\pi$. However, since these indices must be contracted with external legs, they will again induce factors of $t$, either cancelling within themselves due to antisymmetry to give zero contribution (as in our case) or with the poles to yield non-zero but pole-free amplitudes. Similar arguments hold for the absence of $s$- and $u$-channel poles.}
Noting that the symmetry factor for the $s$-channel is again 32, we find
\begin{equation}
i{\cal M}_{\dot{\pi}(\partial\pi)^2,c} = -\frac{1}{2}i \cdot 32 \cdot \frac{(-\alpha_2)^2}{\Lambda^4}\Big[\omega \hskip 2pt \tilde p_1\cdot(\tilde p_1+\tilde p_2)\Big]\,\frac{1}{s}\,\Big[(\tilde p_3+\tilde p_4)\cdot \tilde p_3 \hskip 2pt \omega\Big] = -i\, \alpha_2^2\, \frac{s^2}{\Lambda^4}\ .
\end{equation}
\end{itemize}
\item $\boldsymbol{\dot{\pi}^3\times \dot{\pi}(\partial\pi)^2}$. Finally, we consider the exchange diagram involving both interactions, $\dot \pi^3$ and $\dot{\pi}(\partial\pi)^2$.
There are two such cross-terms, each with amplitude proportional to $\alpha_1 \alpha_2/\Lambda^4$. We have two types of internal contractions:
\begin{itemize}
\item $\contraction{}{\dot{\pi}}{}{\dot{\pi}}\dot{\pi}\dot{\pi}$. There are three ways of obtaining this internal contraction, giving the symmetry factor of $3\times 8=24$ for the $s$-channel. We therefore have
\begin{equation}
i{\cal M}_{\dot{\pi}^3\times\dot{\pi}(\partial\pi)^2,a} = -i \frac{1}{2} \cdot 2 \cdot 24 \cdot \frac{-\alpha_1\alpha_2}{\Lambda^4} \Big[(\tilde p_1\cdot \tilde p_2) 2\omega\Big]\,\frac{1}{s}\,\Big[2\omega \omega^2\Big] = -3i\, \alpha_1 \alpha_2\, \frac{s^2}{\Lambda^4}\ .
\end{equation}
\item $\contraction{}{\dot{\pi}}{}{\partial\pi}\dot{\pi}\partial\pi$. The number of terms with this internal contraction is $3\times 2=6$, so the symmetry factor is $6\times 8=48$. We get
\begin{equation}
i{\cal M}_{\dot{\pi}^3\times\dot{\pi}(\partial\pi)^2,b} =  -i \frac{1}{2}\cdot 2 \cdot 48 \cdot \frac{-\alpha_1 \alpha_2}{\Lambda^4} \Big[\omega \tilde p_1\cdot (\tilde p_1+\tilde p_2)\Big]\,\frac{1}{s}\,\Big[2\omega \omega^2\Big] = -3i\, \alpha_1 \alpha_2\, \frac{s^2}{\Lambda^4}\ .
\end{equation}
\end{itemize}
\end{itemize}

\subsection*{Contact diagrams}

The Lagrangian at quartic order is
\begin{align}
\tilde{\cal L}_4 &= \frac{1}{\Lambda^4}\left[\beta_1\hskip 1pt \dot{\pi}_c^4 - \beta_2\hskip 1pt \dot{\pi}_c^2(\tilde\partial\pi_c)^2 + \beta_3\hskip 1pt (\tilde\partial\pi_c)^4 \right]\ , \label{L4B}
\end{align}
where the parameters $\beta_i$ are defined in (\ref{equ:betas}).
For each contact diagram, we get an overall factor of $i(-i)^2i^2=i$. 
The three interactions in (\ref{L4B}) lead to the following amplitudes:
\begin{itemize}
\item $\boldsymbol{\dot{\pi}^4}$. In the center-of-mass frame, this quartic interaction has equal contributions from $s$-, $t$- and $u$-channels, and comes with a symmetry factor of 24, giving
\begin{align}
i{\cal M}_{\dot{\pi}^4} = i\cdot 24 \cdot \frac{\beta_1}{\Lambda^4} \omega^4 = \frac{3i}{2}\, \beta_1\,\frac{s^2}{\Lambda^4}\ . 
\end{align}
\item $\boldsymbol{\dot{\pi}^2(\partial\pi)^2}$. For this interaction each channel comes with a symmetry factor of 8, and we get
\begin{align}
i{\cal M}_{\dot{\pi}^2(\partial\pi)^2} &= i \cdot 8 \cdot \frac{-\beta_2}{\Lambda^4} \Big[\omega^2(\tilde p_1\cdot \tilde p_2)+\omega^2(\tilde p_1\cdot \tilde p_3)+\omega^2(\tilde p_1\cdot \tilde p_4)\Big] \nonumber\\[4pt]
&= i\, \beta_2\, \frac{s(s-\tilde t-\tilde u)}{\Lambda^4} = 2i\, \beta_2\, \frac{s^2 }{\Lambda^4} \ ,
\end{align}
where we used the relation $s+\tilde t+ \tilde u=0$ to represent the result in terms of $s$ only.
\item $\boldsymbol{(\partial\pi)^4}$. The symmetry factor for this interaction is again 8 for each channel, giving
\begin{align}
i{\cal M}_{(\partial\pi)^4} &=i \cdot  8 \cdot \frac{\beta_3}{\Lambda^4} \Big[(\tilde p_1\cdot \tilde p_2)(\tilde p_3\cdot \tilde p_4)+(\tilde p_1\cdot \tilde p_3)(\tilde p_2\cdot \tilde p_4)+(\tilde p_1\cdot \tilde p_4)(\tilde p_2\cdot \tilde p_3)\Big] \nonumber\\[4pt]
&= 2i\, \beta_3\, \frac{(s^2+\tilde t^{\hskip 1pt 2}+ \tilde u^2)}{\Lambda^4}\ .
\end{align}
Notice that this is the only amplitude with a non-trivial angular dependence.
\end{itemize}

\subsection*{Total amplitude}

Adding the above results, gives the total amplitude 
\beq
 {\cal M}(s,\tilde t\hskip 1pt) \,=\, \left(- \frac{9}{4} \alpha_1^2 - 4 \alpha_2^2 - 6 \alpha_1 \alpha_2 +\frac{3}{2} \beta_1 +2 \beta_2\right) \frac{s^2}{\Lambda^4} + 2 \beta_3\, \frac{(s^2+\tilde t^{\hskip 1pt 2}+ \tilde u^2)}{\Lambda^4}\ .
\eeq
In the forward limit, $\tilde t \to 0$, we find
\beq
{\cal A}(s) \,=\,  \left(-\frac{9}{4} \alpha_1^2 -4\alpha_2^2 - 6 \alpha_1 \alpha_2 + \frac{3}{2} \beta_1+ 2 \beta_2 + 4 \beta_3\right) \frac{s^2}{\Lambda^4}\ .
\eeq

\newpage
\addcontentsline{toc}{section}{References}
\bibliographystyle{utphys}
\bibliography{SignsAnalyticityV3}

\end{document}